\documentclass{article}
\usepackage[utf8]{inputenc}

\pdfoutput=1
\usepackage{amssymb}
\usepackage{amsmath}
\usepackage[dvips]{graphicx}
\usepackage{setspace}
\usepackage{slashed}
\usepackage{mathtools}
\usepackage{amsfonts}
\usepackage{fancyhdr}
\usepackage{xcolor}
\usepackage{graphicx}
\usepackage{rotating}
\usepackage{comment}
\usepackage{color}
\usepackage{subcaption}
\usepackage[percent]{overpic}
\usepackage{cite}
\usepackage{braket}
\usepackage{moresize}
\usepackage{dsfont}

\definecolor{darkgreen}{rgb}{0,0.5,0}
\definecolor{darkblue}{rgb}{0,0,0.6}
\definecolor{purple}{rgb}{0.4,.2,0.7}

\newcommand{\be}{\begin{equation}}
\newcommand{\ee}{\end{equation}}

\usepackage[colorlinks=true,citecolor=darkgreen,linkcolor=black,urlcolor=purple]{hyperref}

\usepackage{pdfsync}

\makeatletter
\newcommand*{\defeq}{\mathrel{\rlap{%
                     \raisebox{0.3ex}{$\m@th\cdot$}}%
                     \raisebox{-0.3ex}{$\m@th\cdot$}}%
                     =} 
\makeatother

\def\be{\begin{eqnarray}}
\def\ee{\end{eqnarray}}

\newcommand{\bea}{\begin{eqnarray}}
\newcommand{\eea}{\end{eqnarray}}
\def\ben{\begin{equation}}
\def\een{\end{equation}}

     \let\r=v

\def\be{\begin{equation}}
\def\ee{\end{equation}}
\def\ba{\begin{eqnarray}}
\def\ea{\end{eqnarray}}

\def\bal#1\eal{\begin{align}#1\end{align}}
\def\bs#1\es{\begin{split}#1\end{split}}

\allowdisplaybreaks  

\numberwithin{equation}{section}

\def\be{\begin{equation}}
\def\ee{\end{equation}}
\def\ba{\begin{eqnarray}}
\def\ea{\end{eqnarray}}
\def\bal#1\eal{\begin{align}#1\end{align}}

\def\r{\rightarrow}

\def\r{\right}

\usepackage{tikz}
\usetikzlibrary{positioning,arrows}
\usetikzlibrary{decorations.pathmorphing}
\usetikzlibrary{decorations.markings}

\usetikzlibrary{%
  arrows,
  knots,
  calc,
}

\tikzset{
  every path/.style={
    red,
    line width=2pt
  },
  every node/.style={
    transform shape,
    knot crossing,
    inner sep=1.5pt
  }
}

\def \be {\begin{equation}}
\def \ee {\end{equation}}

\usepackage{framed}

\begin{document}

\onehalfspacing


\begin{center}

~
\vskip5mm

{\LARGE  {
Statistics in 3d gravity from knots and links
}}

\vskip7mm
Jeevan Chandra

\vskip5mm

{\textit{ Department of Physics, Cornell University, Ithaca, New York, USA 
}\\ and \\ \textit{ Leinweber Institute for Theoretical Physics and Department of Physics, \\ University of California, Berkeley, USA
}}

\vskip5mm

{\tt jn539@cornell.edu }

\end{center}

\vspace{2mm}

\begin{abstract}

In recent years, there has been remarkable progress in evaluating wormhole amplitudes in 3d Einstein gravity with negative cosmological constant and matching them to statistics of 2d CFT data.
In this work, we compute non-perturbative Gaussian and non-Gaussian gravitational contributions to the OPE statistics using a framework that can systematically generate a class of such non-perturbative effects - \textit{Fragmentation of knots and links by Wilson lines}. We illustrate this idea by constructing multi-boundary wormholes from fragmentation diagrams of prime knots and links with upto five crossings. We discuss fragmentations of hyperbolic knots and links like the figure-eight knot, the three-twist knot and the Whitehead link; and non-hyperbolic ones like the Hopf link, the trefoil knot, the Solomon's knot and the Cinquefoil knot. Using Virasoro TQFT, we show how the partition functions on wormholes constructed from different fragmentations of the same knot or link are closely related. Using these fragmentations, we compute gravitational contributions to the variance, a two-point non-Gaussianity, two structures of four-point non-Gaussianities called the `pillow contraction' and the `$6j$-contraction', and some six-point non-Gaussianities. We also check the consistency of some of these non-Gaussianities with the extended Gaussian ensemble of OPE data that incorporates the Gaussian corrections to the variance from knots.

\end{abstract}

\pagebreak

\tableofcontents

\section{Introduction}

Three-dimensional Einstein gravity with negative cosmological constant has been a topic of active research in recent years. Exciting progress has been made toward computing higher topology contributions called Euclidean wormholes to the gravitational path integral and matching the gravitational results holographically to the formal averages of 2d CFT data \cite{Chandra:2022bqq, Belin:2020hea, Collier_2023, Cotler:2020ugk, Schlenker:2022dyo, Jafferis:2024jkb, deBoer:2023vsm}. Different frameworks for computing on-shell and off-shell wormhole amplitudes have been proposed. Some of the important recent works describing off-shell wormhole amplitudes with torus boundaries or Seifert manifolds and their relation to spectral statistics of CFT data include \cite{Cotler:2018zff, Cotler:2020ugk, Maxfield:2020ale, Eberhardt:2022wlc, DiUbaldo:2023qli, Collier:2023cyw, Yan:2023rjh, Haehl:2023mhf, Haehl:2023tkr, Haehl:2023xys, Yan:2025usw, Boruch:2025ilr, deBoer:2025rct}. Semiclassical methods for constructing wormhole solutions to Einstein's equations and computing on-shell actions using the metric formalism have been explored for instance in \cite{Maldacena:2004rf, Chandra:2022bqq, Chandra:2024vhm, Chandra:2024bqz, Chandra:2022fwi, Abajian:2023bqv, Sasieta:2022ksu, Wang:2025bcx}. A TQFT-based framework called Virasoro TQFT was proposed in \cite{Collier_2023} which describes the quantization of 3d gravity on any hyperbolic 3-manifold. Since then, TQFT-based techniques have proven to be quite efficient in computing the gravitational path integral on hyperbolic 3-manifolds exactly in $G_N$ and have been used to compute on-shell wormhole amplitudes, for example, in \cite{Collier:2024mgv, deBoer:2024kat, Post:2024itb, Hartman:2025ula}. These wormholes have been shown to capture the statistics of the OPE data of 2d CFTs consistent with the universal expressions for the moments derived using conformal bootstrap and quantum hyperbolic geometry in \cite{Collier:2018exn, Collier:2019weq, Belin:2021ryy, Anous:2021caj, Ponsot:1999uf, Ponsot:2000mt, Teschner:2012em, Teschner:2013tqy, Verlinde:1989ua, Zamolodchikov:1995aa, Zamolodchikov:2001ah, Dorn:1994xn, Moore:1988qv}. These results help establish that 3d gravity obeys a version of the Eigenstate Thermalization Hypothesis \cite{Srednicki_1994} consistent with Virasoro symmetry called Virasoro ETH proposed in \cite{Collier:2019weq, Belin:2020hea}. A precision test for this realization of ETH was proposed in \cite{Chandra:2024bqz} by constructing wormholes described locally by domain wall solutions, joining CFTs with different couplings. See also \cite{Chandra:2023dgq, Chandra:2023rhx} for some information-theoretic applications of Virasoro ETH. A matrix-tensor model approach unifying the on-shell and off-shell computations, thereby capturing both the spectral and OPE statistics of 2d CFTs, has been proposed in \cite{Belin:2023efa, Jafferis:2024jkb}. This model inspired by the duality between two-dimensional JT gravity and Random Matrix Theory \cite{Saad:2019lba, Saad:2019pqd, Jafferis:2022wez} posits that the formal averages of CFT data described in earlier works can actually be realized as an ensemble of approximate CFTs. There is also a proposal made to sum over topologies using this model in \cite{Jafferis:2024jkb} which would be interesting to make more precise as it is an important outstanding problem in the context of 3d gravity.

The utility of Virasoro TQFT goes beyond holography as it can also be used to compute the partition functions on hyperbolic 3-manifolds which have no asymptotic boundaries. An illustration of this idea was provided in \cite{Collier:2024mgv} where the authors computed the partition function of the complement of the figure-eight knot in $S^3$ (known to be a hyperbolic 3-manifold\footnote{The classic reference on this subject is Thurston's famous lecture notes \cite{Thurston1997}. We also refer the interested reader to \cite{purcell2020hyperbolicknottheory} for a pedagogical modern review of the various mathematical methods involved in the study of hyperbolic knots.}) using Virasoro TQFT, and in the semiclassical limit, they matched it with the known expression for the volume of the 3-manifold thereby verifying the volume conjecture \cite{kashaev1996hyperbolicvolumeknotsquantum} for this case. In this paper, we discuss `fragmentations' of knots (single component links) and two-component links by adding external Wilson lines. As is familiar from the Chern-Simons theory literature \cite{Witten:1988hf}, a knot is a Wilson loop labelled by a conformal weight (usually set to the value at the cusp $\Delta_0=\frac{Q^2}{4}$). Upon adding an external Wilson line, the knot `fragments' into two pieces each described by a different conformal weight. By adding more Wilson lines, we can fragment the knot into more pieces. We can also describe fragmentations of multi-component links in a similar way. Fragmentations allow us to efficiently compute gravitational contributions to the OPE statistics using knots and links. Perhaps the simplest example of this idea of adding Wilson lines to links has already been discussed in \cite{Collier:2024mgv} where they added a Wilson line joining the components of the Hopf link to construct a two-boundary wormhole which computes a two-point non-Gaussianity in the OPE statistics, reviewed in section \ref{twopointnongaussianity} of this paper. 

\subsection{Summary of results}

In this paper, we compute the partition functions on knot and link fragmentations using Virasoro TQFT \cite{Collier_2023} and use the relation between the gravitational partition function and the VTQFT partition function on a hyperbolic 3-manifold $M$,
\begin{equation}
    Z_{\text{grav}}(M)=|Z_V(M)|^2
\end{equation}
to compute the exact partition functions on corresponding multi-boundary wormholes. Although we state our results in terms of wormhole amplitudes with thrice-punctured sphere boundaries so that we can get rid of the unimportant conformal block factors, we can equally well state them as non-perturbative corrections to thermal correlation functions or correlation functions on higher genus Riemann surfaces. In addition, we state our results assuming that all the operators (could be spinning) are above the black hole threshold ($\Delta, \overline{\Delta}>\frac{Q^2}{4}=\frac{c-1}{24}$) just to avoid subtleties associated with analytic continuation of Virasoro crossing kernels to weights below the threshold. Before we summarise our results, we introduce some notation for the Virasoro crossing kernels\footnote{Remarkably, the crossing kernels have been written down in closed form in \cite{Ponsot:1999uf, Ponsot:2000mt, Teschner:2012em, Teschner:2013tqy}. See \cite{Eberhardt:2023mrq} for a comprehensive modern review.} that we shall employ to report all our results,

\begin{itemize}
    \item Liouville structure constants: $C_0(P_1,P_2,P_3) \to C_{123}$.
    \item Fusion kernel: $\mathbb{F}_{P_s,P_t}\begin{bmatrix}
        P_1 & P_2 \\
        P_3 & P_4
    \end{bmatrix} \to \mathbb{F}_{st}\begin{bmatrix}
         1 & 2\\
        3 & 4
    \end{bmatrix} $.
    \item Modular-$\mathbb{S}$ kernel: $\mathbb{S}_{P_a,P_b}[P_c] \to \mathbb{S}_{ab}[c]$.
    \item Braiding phase: $\mathbb{B}^{P_1,P_2}_{P_3}\to \mathbb{B}^{12}_3\equiv e^{i\pi(\Delta_3-\Delta_1-\Delta_2)}$.
    \item Virasoro $6j$-symbol: $\begin{Bmatrix}
        P_1 & P_2 & P_3\\
        P_4 & P_5 & P_6
    \end{Bmatrix} \to \begin{Bmatrix}
       1 & 2 & 3\\
       4 & 5 & 6
    \end{Bmatrix}$
\end{itemize}

In contrast to the Liouville structure constants, we use lowercase $c$ to denote the OPE coefficients between Virasoro primary operators $c_{ijk}\equiv \langle \mathcal{O}_i\mathcal{O}_j\mathcal{O}_k\rangle$. Recall the reality property of OPE coefficients $c_{ikj}=c_{ijk}^*=(-1)^{\ell_i+\ell_j+\ell_k}c_{ijk}$. We will be using this property several times in this paper.
We make a note of some of the important VTQFT identities and the Moore-Seiberg consistency conditions between crossing kernels \cite{Moore:1988qv} that are used in this paper in Appendix \ref{VTQFT identities}. Now, we summarise the results of this paper using the above notation.

\subsubsection*{Variance $\overline{|c_{12a}|^2}$:}

It is well known that the leading contribution\footnote{In this work, whenever we use the term leading contribution to some OPE contraction, we mean the gravitational contribution with minimum number of crossings between Wilson lines in the bulk. It may not be the dominant contribution to the gravitational path integral in all parameter regimes.} is given by the square of the Liouville structure constant \cite{Chandra:2022bqq, Collier_2023},
\begin{equation}
    \overline{|c_{12a}|^2} \supset Z_{\text{grav}}\left[\vcenter{\hbox{

    }} \right]= |C_{12a}|^2
\end{equation}
In section \ref{variance} of this paper, we compute non-perturbative corrections coming from wormholes where two of the worldlines are tangled in the bulk. We construct these wormholes from fragmentations of knots (hyperbolic or non-hyperbolic) by a Wilson line.
The contribution from the two fragmentations of the trefoil knot (a non-hyperbolic knot with three crossings) by a Wilson line are identical and can be collectively expressed as
\begin{equation}
    \overline{|c_{12a}|^2}\supset (\text{phase})|C_{12a}|^2\left |\int dP \rho_0(P)e^{3\pi i P^2} 
\right|^2
\end{equation}
The overall phase depends on the spins of the external operators which is equal to $1$ if we restrict to scalar operators. A wormhole constructed from one of these fragmentations is sketched below,
\begin{align}
    Z_{\text{grav}}\left[\vcenter{\hbox{
    %
\right|^2
\end{align}
We provide details on how to compute the partition functions on these fragmentations of the trefoil knot in section \ref{Trefoiltwopoint}.

We also compute the contributions to the variance coming from fragmentations of the figure-eight knot (a hyperbolic knot with four crossings) in section \ref{Figureeighttwopoint}. The contributions from each of these fragmentations can be expressed as
\begin{equation}
    \overline{|c_{12a}|^2} \supset |C_{12a}|^2 \left |\int dP_s dP_t \rho_0(P_s) \rho_0(P_t) e^{2\pi i (P_s^2-P_t^2)}%
\right|^2
\end{equation}
The wormhole constructed from one of these fragmentations is sketched below,
\begin{multline}
     Z_{\text{grav}}\left[\vcenter{\hbox{
    %
\right|^2
\end{multline}

We also compute the contribution to the variance from fragmentations of the three-twist knot (a hyperbolic knot with five crossings) in Appendix \ref{Three-twist} and from the Cinquefoil knot (a non-hyperbolic knot with five crossings) in Appendix \ref{Cinquefoil}. These two examples are interesting because they allow us to add a Wilson line stretching across more than one crossing of the knot, unlike in the case of the trefoil and figure-eight knots.

At the time when a draft of this work was being written, we came across a talk \cite{TomSCGP} where a similar idea of adding a Wilson line to a hyperbolic knot to compute Gaussian corrections to variance was being discussed.

\subsubsection*{A two-point non-Gaussianity $\overline{c_{11a}c_{22a}^*}$:}

Perhaps the simplest non-Gaussian contraction is $\overline{c_{11a}c_{22a}^*}$. In section \ref{twopointnongaussianity}, we compute gravitational contributions to this non-Gaussianity using fragmentations of two-component links by a Wilson line. The leading contribution comes from the Hopf link and was shown in \cite{Collier:2024mgv} to evaluate to the square of the modular-$\mathbb{S}$ matrix,
\begin{equation}
    \overline{c_{11a}c_{22a}^*}\supset \left|\frac{C_{11a}\mathbb{S}_{12}[a]}{\rho_0(P_2)} \right|^2
\end{equation}
In this paper, we compute non-perturbative corrections from fragmentations of the Solomon's knot (non-hyperbolic link with four crossings), 
\begin{equation}
    \overline{c_{11a}c_{22a}^*}\supset (1+(-1)^{\ell_a})\left|\sqrt{C_{11a}C_{22a}}\int dP \rho_0(P) e^{4\pi i P^2}
 \right|^2
\end{equation}
The wormhole constructed from one of these fragmentations is sketched below,
\begin{equation}
     Z_{\text{grav}}\left[\vcenter{\hbox{
    %
 \right|^2
\end{equation}

We also compute the contributions from the fragmentations of the Whitehead link (a hyperbolic link with five crossings) to this non-Gaussianity,
\begin{equation}
    \overline{c_{11a}c_{22a}^*}\supset (1+(-1)^{\ell_a})\left|\sqrt{C_{11a}C_{22a}}\int dP_sdP_t \rho_0(P_s)\rho_0(P_t)e^{3\pi i(P_s^2+P_t^2)}%
\right|^2
\end{equation}

\subsubsection*{The Pillow contraction $\overline{c_{12a}c_{13a}c_{24b}c_{34b}}$:}

The leading contribution comes from the following 4-boundary wormhole constructed from a fragmentation of the Hopf link by two Wilson lines,
\begin{equation}
    Z_{\text{grav}}\left[ %
\right |^2 
\end{equation}
The wormhole amplitude can also be expressed as a product of R-matrices as shown in (\ref{HopfR}) which makes it convenient to match with the prediction from the Gaussian ensemble of CFT$_2$ data. We also compute the contributions to the pillow contraction from fragmentations of the trefoil knot, figure-eight knot and the Solomon's knot in \ref{fourpointnongaussianities}, and the three-twist knot in Appendix \ref{Three-twist}. We also show how the gravitational results for the pillow contraction are consistent with the CFT$_2$ ensemble that incorporates the Gaussian corrections to variance and the two-point non-Gaussianities discussed earlier.

\subsubsection*{The $6j$-contraction $\overline{c_{12a}c_{34a}c_{23b}c_{41b}}$:}

The leading contribution is given by the square of the $6j$-symbol and is computed by the wormhole discussed in \cite{Collier:2024mgv}. In section \ref{fourpointnongaussianities}, we compute contributions to the $6j$-contraction from fragmentations of the trefoil knot and the figure-eight knot, and in Appendix \ref{Three-twist}, we discuss a fragmentation of the three-twist knot. Analogous fragmentations of two-component links do not contribute to the $6j$-contraction which is unlike the case with the pillow contraction where both knots and two-component links contribute. Below, we have shown a wormhole constructed from fragmentation of the figure-eight knot,
\begin{align}
\begin{split}
   & Z_{\text{grav}}\left[ 
\bigg|^2
    \end{split}
\end{align}

Finally, in section \ref{sixpointnongaussianity}, we compute some structures of six-point non-gaussianities from fragmentations of the Hopf link and the trefoil knot. The Hopf link contribution turns out to be the leading contribution to some of these structures.

\section{Gaussian corrections to variance} \label{variance}

Adding a Wilson line to a knot gives a Gaussian correction to the OPE statistics. As a trivial example, notice that the leading contribution to the variance given by the Liouville structure constant $C_0$ can be expressed in terms of the VTQFT partition function on the unknot with a Wilson line,
\begin{equation}
        Z_V\left [\vcenter{\hbox{

    }} \right]= |C_{123}|^2
    \end{equation}
In this section, we compute Gaussian corrections to the variance from fragmentations of the trefoil knot and the figure-eight knot by one Wilson line, and in Appendices \ref{Three-twist} and \ref{Cinquefoil}, we compute respectively the contributions from the fragmentations of the three-twist knot and the Cinquefoil knots. For the trefoil and figure-eight knot examples, we show that the contribution to the variance from any fragmentation of the knot is identical because the external Wilson line can only stretch across one crossing of the knot. But, for the three-twist and cinquefoil knot examples, we will see that we get different answers for the variance depending on whether the external Wilson line stretches across one or two crossings of the knot.

\subsection{The trefoil knot (3 crossings)} \label{Trefoiltwopoint}

The simplest non-trivial knot is the trefoil knot. In the Alexander-Briggs notation, it is referred to as $3_1$,
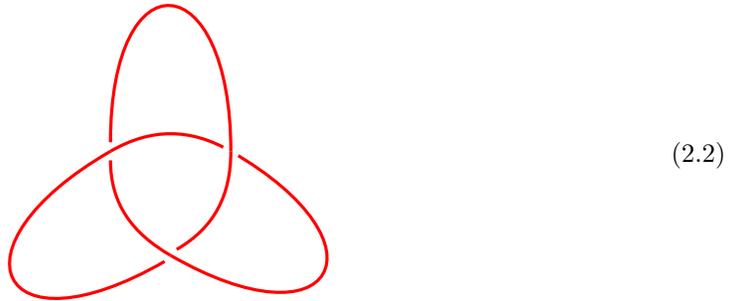
\begin{equation}
    \vcenter{\hbox{
        \begin{tikzpicture}[scale=0.8]
          \draw[very thick, out=-30, in =-30, looseness=4, red] (1,0) to (0,-1.732);
           \draw[fill=white, draw=white] (-1,0) circle (1/10);
           \draw[very thick, out=30, in =150, looseness=1, red] (-1,0) to (1,0);
           \draw[very thick, out=-150, in =-150, looseness=4, red] (-1,0) to (0-0.1,-1.732-0.1);
            \draw[fill=white, draw=white] (1,0) circle (1/10);
             \draw[very thick, out=90, in =90, looseness=4, red] (-1,0.15) to (1,0);
              \draw[very thick, out=-90, in =30, looseness=1, red] (1,0) to (0.1,-1.732+0.1);
              \draw[very thick, out=-90, in =150, looseness=1, red] (-1,-0.15) to (0,-1.732);
        \end{tikzpicture}}}
\end{equation}
It is described by a single Wilson loop with 3 crossings. The complement of the trefoil knot is not hyperbolic so computing the partition function of the $3_1$ knot complement using VTQFT would be ill-defined. However, the addition of a Wilson line facilitates the computation of the partition function using VTQFT. The trefoil knot is not amphichiral i.e, cannot be continuously deformed into its mirror image, so there is a left-handed and right-handed trefoil knot. For concreteness, we work with the right-handed trefoil knot in the explicit computations but the results can be trivially generalized to the left-handed case by flipping the braiding phases.

\begin{figure}
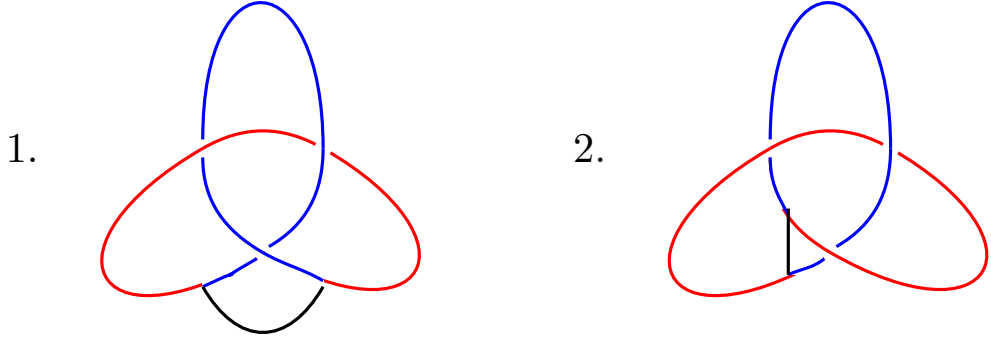

    \centering
    \begin{equation*}
    \vcenter{\hbox{
}}
\end{equation*}
    \caption{The figure above shows the two fragmentations of the right-handed trefoil knot by an external Wilson line drawn in black. The two fragments drawn in blue and red are Wilson lines with different conformal weights. In Fragmentation 1, the blue line has a crossing with itself once and has two crossings with the red line. So, we refer to this fragmentation as `2 cross + 1 self'. In Fragmentation 2, the blue and red lines cross each other thrice. So, we refer to this fragmentation as `3 cross'.}
    \label{fig:Trefoil fragmentations}
\end{figure}

There are two distinct fragmentations for each chirality of the trefoil knot by an external Wilson line as illustrated in figure \ref{fig:Trefoil fragmentations}. We compute below the VTQFT partition functions on both of these fragmentations which in turn gives the gravitational partition function on corresponding two-boundary wormholes. We will observe that the partition functions are given by very similar integral expressions over the same fusion kernel and differ only in an overall braiding phase. However, the contributions to the variance from the two fragmentations are identical and can be collectively written as
\begin{equation}
    \overline{|c_{12a}|^2}\supset (-1)^{\ell_a}|C_{12a}|^2\left |\int dP \rho_0(P)e^{3\pi i P^2} 
\right|^2
\end{equation}
where $\ell_a$ is the spin of operator $\mathcal{O}_a$ assumed to be an integer. Using the hexagon identity, there is an alternate way to express the above result which turns out to be useful for the discussion of non-Gaussianities in section \ref{fourpointnongaussianities},
\begin{equation} \label{trefoilvar}
    \overline{|c_{12a}|^2}\supset |C_{12a}|^2\left |\int dP dP_d\rho_0(P)\rho_0(P_d)e^{i\pi(2 P^2-P_d^2)} %
\right|^2
\end{equation}

\subsubsection{Fragmentation 1 (2 cross + 1 self)}

\begin{align}
    \begin{split}
        Z_V\left[\vcenter{\hbox{
        %
}
    \end{split}
\end{align}
In the first line, we applied $\mathbb{F}$ to the identity line; in the second line, we resolved two of the crossings thereby giving the two braiding phases; we finally evaluated the resulting tetrahedral diagram to give the R-matrix since the lines $1$ and $2$ cross. 

The two-boundary wormhole constructed from this fragmentation of the trefoil knot by embedding the knot diagram into $S^3$ and excising balls around each junction is
\begin{align} \label{trefworm}
    Z_{\text{grav}}\left[\vcenter{\hbox{
    %
\right|^2
\end{align}
Some comments are in order about this expression. Note that the contribution from the left-handed trefoil knot takes the same form with the phase in the integral flipped i.e., $e^{3\pi i \Delta_P}\to e^{-3\pi i \Delta_P}$. When the external Wilson line is removed $P_a \to \frac{iQ}{2}$ so that we just have the undressed trefoil knot with momentum $P_0$, the integral in the above expression reduces to $\int dP\rho_0(P) e^{3\pi i P^2}$. In the semiclassical limit, such an integral does not admit a sensible saddle point thereby reinforcing the non-hyperbolicity of the complement of the trefoil knot. Using the large-$\text{Re}(P)$ asymptotics of the fusion kernel, we can show that that the integral over $P$ does not converge along the real axis. It has to be deformed slightly into the complex plane. We explain this quantitatively in the discussion following the Solomon's knot example in section \ref{twopointnongaussianity} where we end up with a very similar integral. 

\subsubsection{Fragmentation 2 (3 cross)}

\begin{align}
    \begin{split}
         Z_V\left[\vcenter{\hbox{

    \end{split}
\end{align}
The two-boundary wormhole constructed from this fragmentation is
\begin{align}
\begin{split}
    Z_{\text{grav}}\left[\vcenter{\hbox{
    %
\right|^2
    \end{split}
\end{align}
In this diagram, the region shaded in gray is $S^3$, the disks removed (shown in white) are balls in $S^3$ with their boundaries being $\Sigma_{0,3}$.
Noting the cyclic ordering of operator insertions on the two boundaries, we see that the wormhole partition function written above is a contribution to $\overline{c_{12a}^2}$. However, following the standard convention, if we express the result as a contribution to $\overline{|c_{12a}|^2}$, we get,
\begin{equation}
    \overline{|c_{12a}|^2} \supset (-1)^{\ell_a}|C_{12a}|^2\left |\int dP \rho_0(P)e^{3\pi i P^2} \begin{Bmatrix}
            1 & 2 & P \\
            1 & 2 & a
        \end{Bmatrix}\right|^2
\end{equation}

\subsection{The figure-eight knot (4 crossings)} \label{Figureeighttwopoint}

The figure-eight knot is an amphichiral knot with 4 crossings denoted $4_1$ in the Alexander-Briggs notation,
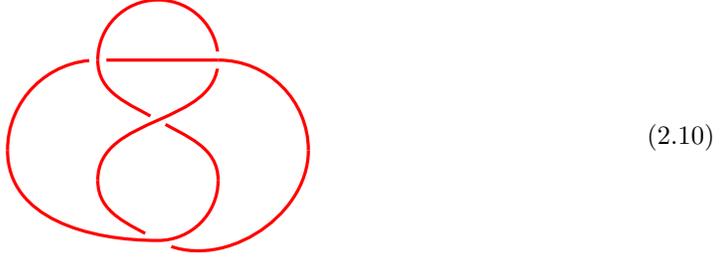
\begin{equation}
   \vcenter{\hbox{ \begin{tikzpicture} [scale=0.8]      
        \draw[very thick, out = 90, in = 180, red] (-5/2,-3/2) to (-1,0);
        \draw[very thick, out = 0, in = 180, red] (-1,0) to (0,0);
        \draw[fill=white,draw=white] (-1,0) circle (1/10);
        \draw[very thick, out = 180, in = 90, red] (0,1) to (-1,0);
\draw[very thick, out = 270, in = 150, red] (-1,0) to (0,-1);
        \draw[very thick, out = 330, in = 90, red] (0,-1) to (1,-2);
        \draw[fill=white,draw=white] (0,-1) circle (1/10);
\draw[very thick, out = 90, in = 270, red] (-1,-2) to (1,0);
        \draw[very thick, out = 90, in = 0, red] (1,0) to (0,1);
        \draw[fill=white,draw=white] (1,0) circle (1/10);
        \draw[very thick, out = 0, in = 180, red] (0,0) to (1,0);
        \draw[very thick, out = 0, in = 90, red] (1,0) to (5/2,-3/2);
        \draw[very thick, out = 270, in = 330, red] (5/2,-3/2) to (0,-3);
\draw[very thick, out = 150, in = 270, red] (0,-3) to (-1,-2);
        \draw[fill=white,draw=white] (0,-3) circle (1/5);
\draw[very thick, out = 270, in= 0, red] (1,-2) to (0,-3);
        \draw[very thick, out = 180, in = 270, red] (0,-3) to (-5/2,-3/2);
\end{tikzpicture}}}
\end{equation}
Its complement in $S^3$ is a hyperbolic 3-manifold with volume 2.02988 (upto 5 decimal places). Its partition function was computed using VTQFT in \cite{Collier:2024mgv} and consistency with the volume conjecture was checked. See also \cite{Dimofte:2009yn, EllegaardAndersen:2011vps, Dimofte:2011gm, Dijkgraaf:2010ur} for corresponding results derived using the Teichmuller TQFT.

\begin{figure}
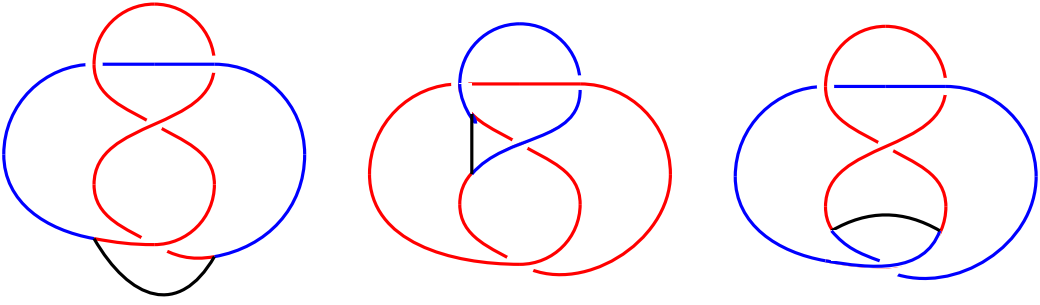

    \centering
    \begin{equation*}
        \vcenter{\hbox{ %
}}
    \end{equation*}
    \caption{The figure above shows the three types of fragmentations of the figure-eight knot by an external Wilson line (shown in black). In the first fragmentation, one of the Wilson line fragments (shown in red) crosses itself twice, and crosses the other fragment (shown in blue) twice. So, we call this fragmentation as `2 cross + 2 self'. Following this terminology, we call the second fragmentation as `3 cross + 1 self' and the third fragmentation as `2 cross + (1 + 1) self'. In addition, there is are `mirror' fragmentations to each of these with the pattern of over- and under- crossings of the Wilson lines attached to the external Wilson line reversed.}
    \label{fig:Figure-eight fragmentations}
\end{figure}

Now, we discuss the fragmentations of the figure-eight knot and their relation to the variance of OPE coefficients. We also reproduce the VTQFT partition function on the figure-eight knot computed in \cite{Collier:2024mgv} below by taking an identity limit of the external Wilson line.
The fragmentations of the figure-eight knot by an external Wilson line are shown in figure \ref{fig:Figure-eight fragmentations} with a description in the caption. We compute the gravitational partition functions on the wormholes constructed from these fragmentations in the subsequent part of the section. The contributions from these wormholes toward the variance $\overline{|c_{12a}|^2}$ are identical and can be written as,
\begin{equation} \label{fig8var}
    \overline{|c_{12a}|^2} \supset |C_{12a}|^2 \left |\int dP_s dP_t \rho_0(P_s) \rho_0(P_t) e^{2\pi i (P_s^2-P_t^2)}
\right|^2
\end{equation}
Note as a consistency check that upon removing the phase in the integral, the integral evaluates to $1$ owing to the idempotency of the $6j$-symbol.

\subsubsection{Fragmentation 1 (2 cross+ 2 self)}

\begin{align} \label{Frag1fig8}
    \begin{split}
        Z_V\left[\vcenter{\hbox{ 

    \end{split}
\end{align}
The gravitational partition function on the corresponding two-boundary wormhole is given by
\begin{multline} \label{gravfig8}
     Z_{\text{grav}}\left[\vcenter{\hbox{
    %
\right|^2
\end{multline}
 When expressed as a contribution to the variance $\overline{|c_{12a}|^2}$,
 \begin{equation}
     \overline{|c_{12a}|^2} \supset |C_{12a}|^2\left |\int dP_s dP_t \rho_0(P_s) \rho_0(P_t) e^{2\pi i (P_s^2-P_t^2)}%
\right|^2
 \end{equation}
Some comments are in order. It is easy to check that in the identity limit $P_a\to \frac{iQ}{2}$, the above VTQFT expression reduces to the partition function of the figure-eight knot with Liouville momentum $P_0$,
\begin{equation}
    Z_V[4_1]=\int dP_s dP_t \rho_0(P_s) \rho_0(P_t) e^{2\pi i (P_s^2-P_t^2)}
     %
\end{equation}
The above expression for the partition function of the figure-eight knot is unchanged when the phase inside the integral is flipped. This is expected since the figure-eight knot is amphichiral. However, when we consider the partition function with the external Wilson line (\ref{Frag1fig8}), upon flipping the braiding phase, the expression is no longer the same. But note that we can get the expression with the opposite braiding phase by adding the external Wilson line at a different location on the figure-eight knot. So there is an imprint of amphichirality even with the addition of an external Wilson line. Also note that with the removal of the braiding phase in the integral expression for the gravitational partition function in (\ref{gravfig8}), using the idempotency of the $6j$-symbol, we recover the expected result of $|C_{12a}|^2$ for the unknotted wormhole. 

\subsubsection{Fragmentation 2 (3 cross + 1 self)}

We can similarly compute the partition function on fragmentation $2$. We skip the details and just present the result,
\begin{align}
    \begin{split}
        & Z_V\left[\vcenter{\hbox{ 

    \end{split}
\end{align}
To arrive at this expression, we first resolve all the crossings and compute the partition function of the resulting hyperbolic tetrahedron just like in the previous fragmentation. To arrive at the answer quickly, we could just use the VTQFT partition function on the subdiagram (\ref{crossvert}) in Appendix \ref{Useful identities}.

The two-boundary wormhole constructed from this fragmentation is shown below,
\begin{align}
\begin{split}
     & Z_{\text{grav}}\left[\vcenter{\hbox{
    %
\right |^2
     \end{split}
\end{align}
In this diagram, the region shaded in gray is $S^3$, the disks removed (shown in white) are balls in $S^3$ with their boundaries being $\Sigma_{0,3}$.
When interpreted as a contribution to $\overline{|c_{12a}|^2}$, we can get rid of the prefactor thereby giving
\begin{equation}
    \overline{|c_{12a}|^2} \supset |C_{12a}|^2 \left |\int dP_s dP_t \rho_0(P_s) \rho_0(P_t) e^{2\pi i (P_s^2-P_t^2)}%
\right |^2
\end{equation}

\subsubsection{Fragmentation 3 (2 cross + (1+1) self)}

The computation of the partition function of fragmentation 3 proceeds in exactly the same way as that of fragmentation 1 so we skip the details and state the result,
\begin{equation} \label{frag3fig8}
    Z_V\left[\vcenter{\hbox{ %
\end{equation}
Comparing the above expression with the results of the other fragmentations, we see that the only difference is in the overall phase, which captures the effect of moving the external Wilson line through the crossing. We have also observed this fact about moving a Wilson line through a crossing more generally in Appendix \ref{Useful identities}.
Since the operators have integer spin, the effect of this phase is removed in the expression for the gravitational partition function of the wormhole,
\begin{multline}
     Z_{\text{grav}}\left[\vcenter{\hbox{
    %
\right|^2
\end{multline}
 When expressed as a contribution to the variance $\overline{|c_{12a}|^2}$,
 \begin{equation}
     \overline{|c_{12a}|^2} \supset |C_{12a}|^2\left |\int dP_s dP_t \rho_0(P_s) \rho_0(P_t) e^{2\pi i (P_s^2-P_t^2)}%
\right|^2
 \end{equation}

\subsubsection{A no-go for all-cross fragmentation}

Note that interestingly, there is no way to fragment the figure-eight knot in such a way that the two fragments cross each other 4 times, unless the external Wilson line crosses the knot fragments which we do not allow in the present discussion. This is unlike the case of the trefoil knot where we found a fragmentation with the two fragments crossing each other 3 times. This no-go result for the all-cross fragmentation suggests that the corresponding two-boundary wormhole where two of the worldlines cross each other 4 times does not exist.

\section{A two-point non-Gaussianity} \label{twopointnongaussianity}

Perhaps the simplest contraction resulting in a non-Gaussianity is $\overline{c_{iik}c_{jjk}^*}$. This two-point non-Gaussianity receives contributions from two-boundary wormholes constructed from two-component links joined by a Wilson line. The simplest non-trivial link is the Hopf link which has two crossings. More non-trivial examples include the Solomon's knot (actually a two-component link) and the Whitehead link with $4$ and $5$ crossings respectively. Below, we discuss the fragmentations of these links that contribute to the two-point non-Gaussianity. 

\subsection{The Hopf link (2 crossings)}

The contribution to the two-point non-Gaussianity coming from the two-boundary wormhole constructed by adding a Wilson line to the Hopf link was calculated in \cite{Collier:2024mgv} in analogy with corresponding results in TQFTs based on modular tensor categories \cite{Kitaev:2005hzj}. Here, we simply state their result for completeness,
\begin{align}
\begin{split}
     Z_V\left[\vcenter{\hbox{

   \\
    =& e^{-i\pi \Delta_a}\frac{\mathbb{S}_{12}[a]}{\rho_0(P_2)C_{22a}}
    \end{split}
\end{align}
with the gravitational contribution given by
\begin{equation} \label{hopftwopoint}
    \overline{c_{11a}c_{22a}^*}\supset Z_{\text{grav}}\left[
    \vcenter{\hbox{
    %
    }}
\right]= \left|\frac{C_{11a}\mathbb{S}_{12}[a]}{\rho_0(P_2)} \right|^2
\end{equation}

\subsection{The Solomon's knot (4 crossings)} \label{Soltwopoint}

The Solomon's knot is a two-component link with 4 crossings sketched below,
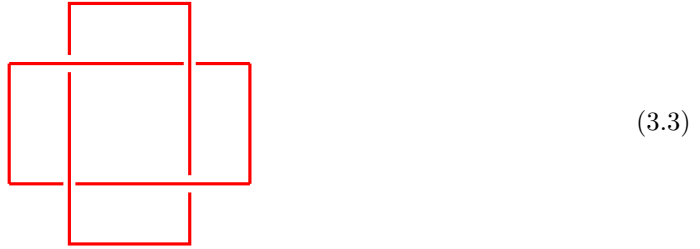
\begin{equation}
     \vcenter{\hbox{%
}}
\end{equation}
Its complement in $S^3$ is not a hyperbolic 3-manifold. However, with the addition of a Wilson line joining the two components, we can compute the partition function on the resulting network using VTQFT.
There are four fragmentations of the Solomon's knot contributing to the two-point non-Gaussianity corresponding to the four different ways in which the external Wilson line can stretch across a crossing between the two components of the link. They are shown in the figure below,
\begin{align}
\begin{split}
    \vcenter{\hbox{%
}}
    \end{split}
\end{align}
The blue and red rectangles are the two fragments labelled by different conformal weights. The black line is the external Wilson line. 
We can easily compute the VTQFT partition functions on these fragmentations by introducing an identity line between the two components of the link and undoing the braidings after fusion on this line. The VTQFT partition functions on fragmentations 3 and 4 are respectively equal to those on fragmentations 1 and 2 since there is no effect of moving both the end points of the external Wilson line across the crossing. See Appendix \ref{Useful identities} for a simple reason. In fact, as we can see from the figure, 1 and 3 are identical fragmentations even without the need to compare their VTQFT partition functions. 

\subsubsection{Fragmentation 1}

The partition function on the first fragmentation is given by
\begin{equation}
    Z_V\left[ \vcenter{\hbox{

\end{equation}
To arrive at the above expression, we applied a $\mathbb{F}$-move on the identity line which suffices to resolve all the crossings thereby giving the braiding phase $\left ( \mathbb{B}^{1,2}_P \right)^4$ and a tetrahedral diagram.
The gravitational partition function on the corresponding 2-boundary wormhole given by 
\begin{equation} \label{solfrag1}
     Z_{\text{grav}}\left[\vcenter{\hbox{
    %
 \right|^2
\end{equation}
As written, the above expression is a gravitational contribution to $\overline{c_{11a}c_{22a}^*}$. We make some comments about the choice of integration contour which are also applicable with appropriate modifications to other examples.
If the integration contour for $P$ is along the positive real axis, then the integral does not converge. To see this, recall the large $\text{Re}(P)$ asymptotics of the $\mathbb{F}$-kernel \cite{Collier:2018exn},
\begin{equation}
    \frac{\mathbb{F}_{aP}%
} \to e^{-2\pi P(\frac{Q}{2}+iP_a)}(\dots)
\end{equation}
where $(\dots)$ is a $P$-independent prefactor. The integrand therefore has the following asymptotics for large $\text{Re}(P)$,
\begin{equation}
    \text{integrand} (P) \sim e^{\pi P Q+4\pi i P^2-2\pi i PP_a}
\end{equation}
Therefore, for the integral to converge, we could choose a contour that starts at $P=0$ and asymptotes to $\mathbb{R}+i\beta$ for large $\text{Re}(P)$ with $\beta>\frac{1}{8}(Q+2\text{Im}(P_a))$. Since $0\leq \text{Im}(P_a)<\frac{Q}{2}$, we may as well choose $\beta>\frac{Q}{4}$.

\subsubsection{Fragmentation 2}

The partition function on the second fragmentation evaluates to \\
\begin{equation}
     Z_V\left[ \vcenter{\hbox{

\end{equation}
with the gravitational partition function on the corresponding 2-boundary wormhole given by 
\begin{align}
\begin{split}
   Z_{\text{grav}}\left[\vcenter{\hbox{
    %
\right|^2
    \end{split} 
    \end{align}
Using the hexagon identity, the above contribution to the two-point non-Gaussianity from this fragmentation can also be expressed as
\begin{equation} \label{Solfrag2}
     \overline{c_{11a}c_{22a}^*}\supset \left|\sqrt{C_{11a}C_{22a}}\int dPdP_d \rho_0(P)\rho_0(P_d) e^{ i\pi(3 P^2-P_d^2)}%
 \right|^2
\end{equation}
We shall find this expression handy in the next section where we discuss four-point non-Gaussianities. It is interesting to note that unlike all the previous examples, the two fragmentations of the Solomon's knot give different contributions to the same OPE contraction. This observation is going to be important in the next section.

The contributions of the two fragmentations can be collectively summarised as
\begin{equation}
    \overline{c_{11a}c_{22a}^*}\supset (1+(-1)^{\ell_a})\left|\sqrt{C_{11a}C_{22a}}\int dP \rho_0(P) e^{ 4\pi i P^2}%
 \right|^2
\end{equation}

\subsection{The Whitehead link (5 crossings)}

The Whitehead link is a two-component link with 5 crossings. In the Alexander-Briggs notation, it is denoted $5_1^2$,
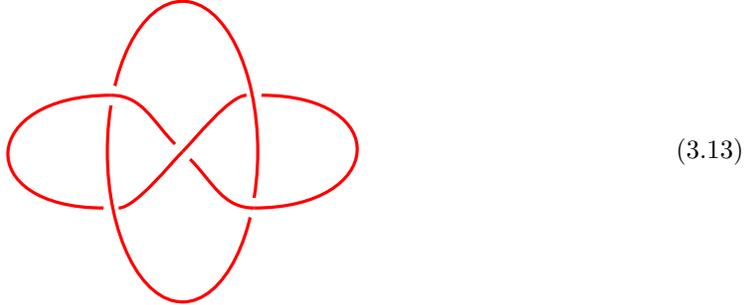
\begin{equation}
     \vcenter{\hbox{%
}}
\end{equation}
It is a chiral link.
Its complement in $S^3$ is a hyperbolic 3-manifold of volume $3.664$ (rounded to 3 decimal places). So, its partition function can be computed using VTQFT. By introducing two identity lines as shown, we can apply fusion and undo the braidings to reduce to the hyperbolic tetrahedron,
\begin{align} \label{whiteheadpartition}
\begin{split}
    Z_V\left[\vcenter{\hbox{%
    \end{split}
    \end{align}
Notice that the partition function takes a similar form to the partition function on the figure-eight knot complement with the difference being the braiding phases inside the integral. Using the known semiclassical expansion of the $6j$-symbol, one could try to verify the volume conjecture for the whitehead link,
\begin{equation}
    |Z_V[5_1^2]|=e^{-\frac{c}{12\pi}\text{Vol}(5_1^2)}
\end{equation}
The modulus is necessary since the whitehead link is chiral. We will not present the details here as it is not important for the present work.

\subsubsection{Fragmentations of the whitehead link}

With the addition of an external Wilson line joining the two components, just like with the Solomon's knot example, there are four fragmentations corresponding to the four different ways in which the external Wilson line stretches across a crossing between the two components of the link as shown below,
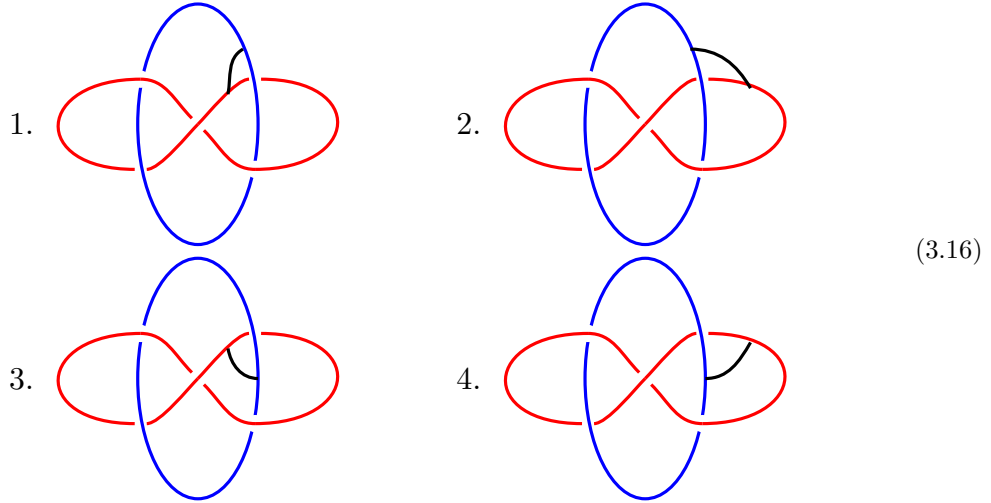
\begin{align}
\begin{split}
    \vcenter{\hbox{
}}
    \end{split}
\end{align}
The VTQFT partition functions on fragmentations 3 and 4 are respectively equal to those on fragmentations 1 and 2 as there is no effect of moving both the end points of the Wilson line across the crossing. So we restrict discussion to fragmentations 1 and 2.
The partition functions on these two fragmentations can be computed in a similar way, so we skip the details and just write down the result. Assigning momentum $P_1$ to the red fragment, momentum $P_2$ to the blue fragment and $P_a$ to the external Wilson line, we have
\begin{equation}
    Z_V\left[\vcenter{\hbox{%
\end{equation}
In the identity limit $P_a \to \frac{iQ}{2}$, we recover the partition function on the whitehead link computed in (\ref{whiteheadpartition}).
The gravitational partition function on the corresponding two-boundary wormhole is given by
\begin{equation}
    \overline{c_{11a}c_{22a}^*} \supset Z_{\text{grav}}=\left|\sqrt{C_{11a}C_{22a}}\int dP_sdP_t \rho_0(P_s)\rho_0(P_t)e^{3\pi i (P_s^2+P_t^2)}%
\right|^2
\end{equation}

Similarly, the VTQFT partition function on the second fragmentation is given by
\begin{equation}
     Z_V\left[\vcenter{\hbox{%
\end{equation}
with the gravitational partition function on the corresponding two-boundary wormhole given by
\begin{equation}
   \overline{c_{11a}c^*_{22a}} \supset Z_{\text{grav}}=(-1)^{\ell_a}\left|\sqrt{C_{11a}C_{22a}}\int dP_sdP_t \rho_0(P_s)\rho_0(P_t)e^{3\pi i(P_s^2+ P_t^2)}%
\right|^2
\end{equation}
Just like with the Solomon's knot, the fragmentations of the Whitehead link give different contributions to the same OPE contraction.
In summary, the contribution from the two fragmentations of the Whitehead link to the two-point non-Gaussianity can be expressed collectively as
\begin{equation}
    \overline{c_{11a}c_{22a}^*}\supset (1+(-1)^{\ell_a})\left|\sqrt{C_{11a}C_{22a}}\int dP_sdP_t \rho_0(P_s)\rho_0(P_t)e^{3\pi i (P_s^2+P_t^2)}%
\right|^2
\end{equation}

\section{Four-point non-Gaussianities} \label{fourpointnongaussianities}

In this section, we discuss the fragmentations of various knot and links by a pair of external Wilson lines. These fragmentations correspond to 4-boundary wormholes which contribute to 4-point non-Gaussianities in the OPE statistics. Specifically, we compute contributions to the two structures of the fourth moment of OPE coefficients between distinct operators, involving six distinct operators in total,
\begin{enumerate}
    \item \textbf{The pillow contraction}: This is the fourth moment given by $\overline{c_{12a}c_{13a}c_{24b}c_{34b}}$. This contraction needs a minimum of two crossings between worldlines in the wormhole to give a non-trivial contribution. Hence, fragmentations of the Hopf link by two Wilson lines gives the leading (minimal crossings) contribution to the pillow contraction. We also compute the contributions coming from fragmentations of the trefoil knot, Solomon's knot and the figure-eight knot in the rest of this section and in Appendix \ref{Three-twist}, we discuss the contribution from the three-twist knot. We also show how the gravitational results for the pillow contraction are consistent with the CFT$_2$ ensemble that incorporates the Gaussian corrections to variance and the two-point non-Gaussianities discussed in the previous sections.
    \item \textbf{The $6j$-contraction}: This is the fourth moment given by $\overline{c_{12a}c_{34a}c_{23b}c_{41b}}$. It is termed the $6j$-contraction because the leading contribution is given by the square of the Virasoro $6j$-symbol,
    \begin{equation}
        \overline{c_{12a}c_{34a}c_{23b}c_{41b}}\supset \left |\sqrt{C_{12a}C_{34a}C_{23b}C_{41b}}\begin{Bmatrix}
            1 & 2 & a\\
            3 & 4 & b
        \end{Bmatrix}\right |^2
    \end{equation}
    It is easy to see that fragmentations of two-component links like the Hopf link and the Solomon's knot do not contribute to the $6j$-contraction. So, we will compute the non-perturbative corrections to the $6j$-symbol coming from the fragmentations of the trefoil knot and the figure-eight knot, and in Appendix \ref{Three-twist}, we compute the contribution from the three-twist knot.
\end{enumerate}

\subsection{The Hopf link (2 crossings)}

In this section, we compute the contribution to the pillow contraction of OPE coefficients $\overline{c_{12a}c_{13a}c_{24b}c_{34b}}$ from the Hopf link fragmented by two Wilson lines. First, we compute the VTQFT partition function on the setup by introducing an identity line between the circles and applying the $\mathbb{F}$-move,
\begin{align} \label{Hopf1}
\begin{split}
    Z_V\left[\vcenter{\hbox{

    \end{split}
\end{align}
Note that without the braiding phase, the idempotency of $\mathbb{F}$ would mean the integral evaluates to the $\delta$ function setting the Liouville momenta of $a$ and $b$ to be the same,
\begin{equation}
    \int dP  \mathbb{F}_{aP}%
=\delta(P_a-P_b)
\end{equation}
The corresponding setup would be an unlink joined by two Wilson lines,
\begin{equation}
     Z_V\left[\vcenter{\hbox{
        %
  }}
    \right]=\frac{\delta(P_a-P_b)}{\rho_0(P_b)C_{12b}C_{34b}}
\end{equation}

We can construct a 4-boundary wormhole from this configuration of Wilson lines by embedding the setup in $S^3$ and excising balls around each junction. Using the correct normalisation of such a junction, we can compute the VTQFT partition function and hence the gravitational partition function on the 4-boundary wormhole which provides the leading contribution to the 4-point non-Gaussianity $\overline{c_{12a}c_{3a4}c_{34b}c_{1b2}}$,
\begin{equation} \label{hopfworm}
    Z_{\text{grav}}\left[ 
\right |^2 
\end{equation}

\subsubsection{Heegaard splitting along twice-punctured tori}

Alternatively, we can compute the partition function by Heegaard splitting along twice-punctured tori. The resulting states on $\Sigma_{1,2}$ obtained by evaluating the VTQFT partition functions on the corresponding generalized compression bodies are
\begin{equation}
    \begin{split}
        & \bigg |\vcenter{\hbox{
        %
}}\bigg \rangle 
    \end{split}
\end{equation}
We want to evaluate the inner product between these states. But since the tori are interlocked, we first do an $\mathbb{S}$-transformation to bring the blocks on the RHS to the same channel and then evaluate the inner product using
\begin{equation}
    \bigg \langle\vcenter{\hbox{
        %
}} \bigg \rangle =\frac{\delta(P-P_4)\delta(P_c-P_d)}{\rho_0(P)\rho_0(P_c)C_{abc}C_{44d}}
\end{equation}
thereby giving an alternate expression for the VTQFT partition function on the Hopf link network,
\begin{equation}
    \begin{split}
         Z_V\left[\vcenter{\hbox{
        %
     \end{split}
\end{equation}
To arrive at the last line, we used the relation between the $\mathbb{S}$-kernel and the $\mathbb{F}$-kernel. Requiring consistency of VTQFT, the two representations should be equivalent. So, we arrive at the following integral identity obeyed by the crossing kernels,
\begin{multline}
    \int dP_c dP \frac{(\mathbb{B}^{2,4}_P)^{2}}{\rho_0(P_c)C_{24P}C_{abc}}\mathbb{F}_{1c}%
\end{multline}
To independently verify that these two expressions are equivalent, note that
\begin{equation}
    \frac{1}{C_{24P}}\int dP_c\frac{1}{\rho_0(P_c)C_{abc}}\mathbb{F}_{1c}%
\end{equation}
thanks to the pentagon identity. To see this more clearly, we can rewrite the above equality as
\begin{equation} \label{penta1}
    \int dP_c \mathbb{F}_{1c}%
\end{equation}
which is the familiar form of the pentagon identity.
Conversely, we could view the above VTQFT calculation as a three-dimensional derivation of the pentagon identity.

\subsubsection{Consistency with the \texorpdfstring{CFT$_2$}{CFT2} ensemble} \label{hopfCFT}

In order to check the consistency of the gravitational result with the CFT$_2$ ensemble \cite{Chandra:2022bqq}, we find it convenient to evaluate the partition function on the Hopf link network in yet another way, in terms of a product of R-matrices,
\begin{align} \label{Hopf3}
    \begin{split}
         Z_V\left[\vcenter{\hbox{

    \end{split}
\end{align}
In the first line, we applied $s-u$ crossing moves on the external Wilson lines to get the product of R-matrices. The gravitational partition function therefore also has an alternate representation in terms of a product of R-matrices,
\begin{equation} \label{HopfR}
    Z_{\text{grav}}\left[ %
\right |^2 
\end{equation}
The advantage of writing the partition function as a product of R-matrices is that it becomes natural to match the result with the prediction from the CFT$_2$ ensemble by expanding the averaged product of two 4-point functions using the u-channel,
\begin{align}
    \begin{split}
\overline{\langle\mathcal{O}_1\mathcal{O}_4\mathcal{O}_2\mathcal{O}_3\rangle\langle\mathcal{O}_1\mathcal{O}_4\mathcal{O}_2\mathcal{O}_3\rangle^*}=&\sum_{a,b}\overline{c_{12a}c_{3a4}c_{34b}c_{1b2}}\left |\,\, \vcenter{\hbox{%
}}  \,\,\quad \right |^2
    \end{split}
\end{align}
In the second line, we substituted the expression for the non-Gaussianity computed from the 4-boundary wormhole (\ref{HopfR}) using which we reproduced the expectation from the Gaussian ensemble in the last line. Also note that requiring consistency between the two VTQFT partition function (\ref{Hopf1}) and (\ref{Hopf3}) gives the following identity obeyed by the crossing kernels,
\begin{align}
\begin{split}
  e^{2\pi i(\Delta_1+\Delta_2+\Delta_3+\Delta_4) } \int dP \rho_0(P) e^{-2\pi i \Delta_P}&%
\end{split}
\end{align}
This identity can also be derived independently using the properties of the $6j$-symbol. To do so, introduce a $\delta$-function on the LHS and using the idempotency of the $6j$-symbol, express it as the integral of a product of $6j$-symbols, and finally apply the hexagon identity twice to get the RHS.

Now, we discuss another interpretation of the wormhole partition function (\ref{hopfworm}) which shows that it is consistent with the CFT$_2$ ensemble that incorporates the two-point non-Gaussianity (\ref{hopftwopoint}). This involves computing the averaged product of the 4-point functions $\overline{\langle \mathcal{O}_a \mathcal{O}_2 \mathcal{O}_2 \mathcal{O}_b \rangle \langle \mathcal{O}_a \mathcal{O}_4 \mathcal{O}_4 \mathcal{O}_b \rangle^*}$. First, we compute the average by expanding both the 4-point functions using the $s$-channel and use the gravitational result (\ref{hopfworm}) for the non-Gaussianity,
\begin{align} \label{hopfschannel}
    \begin{split}
        &\overline{\langle \mathcal{O}_a \mathcal{O}_2 \mathcal{O}_2 \mathcal{O}_b \rangle \langle \mathcal{O}_a \mathcal{O}_4 \mathcal{O}_4 \mathcal{O}_b \rangle^*}\\ \\
        & =\sum_{1,3}\overline{c_{12a}c_{1b2}c_{3a4}c_{34b}}\left|  \vcenter{\hbox{
}}\bigg |^2
    \end{split}
\end{align}
Now, we evaluate the average by expanding both the 4-point functions using the $t$-channel,
\begin{align}
    \begin{split}
        &\overline{\langle \mathcal{O}_a \mathcal{O}_2 \mathcal{O}_2 \mathcal{O}_b \rangle \langle \mathcal{O}_a \mathcal{O}_4 \mathcal{O}_4 \mathcal{O}_b \rangle^*}\\ \\
        & =\sum_{d,d'}\overline{c_{abd}c_{22d}c^*_{abd'}c^*_{44d'}}\left|  \vcenter{\hbox{%
}}\bigg |^2
    \end{split}
\end{align}
In the second line, we used (\ref{hopftwopoint}). In the third line, we expressed the $t$-channel blocks in terms of the $s$-channel blocks. In the last line, we used the pentagon identity to evaluate the integral over $P$,
\begin{align}
    \int dP\rho_0(P) %
\end{align}
This is the same form of the pentagon identity as written in (\ref{penta1}), expressed here in terms of the $6j$-symbol instead of the $\mathbb{F}$-kernel.
Since the $s$-channel and $t$-channel expansions agree, we conclude that the wormhole partition function (\ref{hopfworm}) is consistent with the CFT$_2$ ensemble that incorporates the two-point non-Gaussianity (\ref{hopftwopoint}).

\subsection{The trefoil knot (3 crossings)}

\subsubsection{Contribution to the pillow contraction}

Shown below are the three fragmentations of the trefoil knot corresponding to wormholes which contribute to the pillow contraction of 4 OPE coefficients,
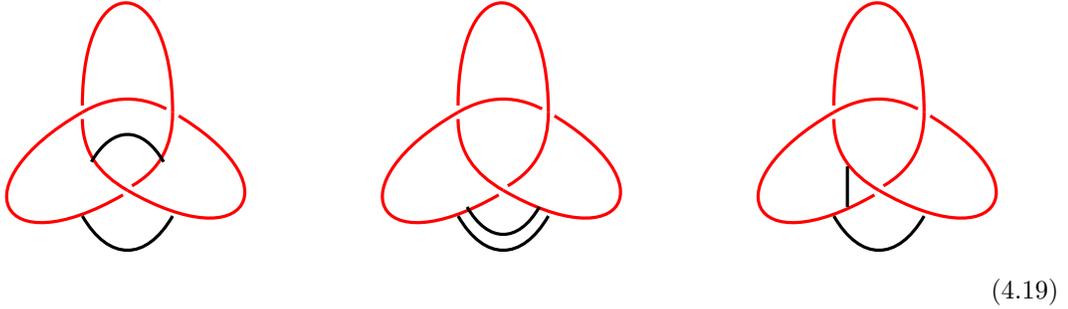
\begin{align} \label{fragtrefoilpill}
     \vcenter{\hbox{
}}
\end{align}
Although the wormholes constructed from these fragmentations are different, their contributions to the pillow contraction are equivalent and given by
\begin{align}
\begin{split}
   \overline{c_{12a}c_{13a}c_{24b}c_{34b}} \supset (\text{phase})\bigg|\sqrt{C_{12a}C_{13a}C_{24b}C_{34b}} \int & dP dP_d \rho_0(P) \rho_0(P_d) e^{i\pi (2P^2-P_d^2)}\\ & \qquad\times %
\bigg |^2
        \end{split}
\end{align}

We now show that the partition functions on these fragmentations only differ by an overall phase,
\begin{align}
    \begin{split}
        Z_V\left[\vcenter{\hbox{
        %
    \end{split}
\end{align}
From the knot diagram, we see that exchanging $P_2$ and $P_3$ should leave the result unchanged. We see that the VTQFT partition function is invariant under $P_2 \xleftrightarrow{} P_3$.
It is also easy to check that the partition function changes by the following phase when the Wilson line is pushed downward through the crossing,
\begin{equation}
    Z_V\left[\vcenter{\hbox{
        %
}}\right]
\end{equation}
This means that the wormholes corresponding to these two diagrams have the same partition function upto an overall phase dependent on the spin of the external operators.
The 4-boundary wormhole corresponding to the diagram on the right is drawn below,
\begin{equation} \label{trefoilwormpillow}
    \overline{c_{12a}c_{1a3}c_{24b}c_{3b4}} \supset Z_{\text{grav}}\left[ %
\bigg |^2
        \end{split}
\end{align}

Finally, we turn to the third fragmentation in (\ref{fragtrefoilpill}). In this diagram, we see that each of the 4 fragments has a different number of crossings. Assigning momentum $P_2$ to the fragment which has zero crossings; $P_3$ to the fragment with one crossing; $P_1$ to the fragment with two crossings; and $P_4$ to the fragment with four crossings, the VTQFT partition function is given by
\begin{align}
    \begin{split}
       & Z_V\left[\vcenter{\hbox{
        %
    \end{split}
\end{align}
which upto the overall phase, agrees with the previous results thus confirming our claim that the fragmentations are equivalent.

Now, we check that the above gravitational result (\ref{gravtrefoilpill}) is consistent with the extended Gaussian ensemble of CFT$_2$ that incorporates the non-perturbative correction to the variance coming from the fragmentations of the trefoil knot computed in section \ref{Trefoiltwopoint}. To this end, we compute the averaged product of two four-point functions $\overline{\langle \mathcal{O}_a\mathcal{O}_1\mathcal{O}_4\mathcal{O}_b \rangle \langle \mathcal{O}_a\mathcal{O}_1\mathcal{O}_4\mathcal{O}_b \rangle^*}$ in two different ways and check that they agree. First, we compute the average by expanding both the four-point functions using the $s$-channel,
\begin{align} \label{trefoilschannel}
    \begin{split}
        &\overline{\langle \mathcal{O}_a\mathcal{O}_1\mathcal{O}_4\mathcal{O}_b \rangle \langle \mathcal{O}_a\mathcal{O}_1\mathcal{O}_4\mathcal{O}_b \rangle^*}\\ \\
        & =\sum_{2,3}\overline{c_{12a}c_{24b}c_{1a3}c_{3b4}}\left|  \vcenter{\hbox{
}} \bigg |^2
    \end{split}
\end{align}
where we substituted the gravity result (\ref{gravtrefoilpill}) for the non-Gaussianity. Note that the above averaged product of 4-point functions is also the gravitational partition function on a 2-boundary wormhole with the boundaries being 4-punctured spheres. The worldlines sourced by operators $\mathcal{O}_a$ and $\mathcal{O}_b$ are unknotted and go straight across the wormhole while the worldlines sourced by operators $\mathcal{O}_1$ and $\mathcal{O}_4$ are knotted into a trefoil knot in exactly the same way as shown in (\ref{trefoilwormpillow}).
Now, we compute the average by expanding both the four-point functions using the $t$-channel,
\begin{align}
    \begin{split}
        &\overline{\langle \mathcal{O}_a\mathcal{O}_1\mathcal{O}_4\mathcal{O}_b \rangle \langle \mathcal{O}_a\mathcal{O}_1\mathcal{O}_4\mathcal{O}_b \rangle^*}\\ \\
        & =\sum_{s,t}\overline{c_{abs}c_{1s4}c^*_{abt}c^*_{1t4}}\left|  \vcenter{\hbox{
}} \bigg|^2
    \end{split}
\end{align}
In the second line, we used (\ref{trefoilvar}) to evaluate the Gaussian average. In the third line, we expressed the $t$-channel blocks in terms of the $s$-channel blocks. In the last line, we used the pentagon identity to evaluate the integral over $P_s$,
\begin{align}
    \int dP_s \rho_0(P_s) %
\end{align}
Since the two ways of computing the average match, we conclude that the gravitational result (\ref{gravtrefoilpill}) is consistent with the extended Gaussian ensemble that incorporates the correction (\ref{trefoilvar}).

\subsubsection{Contribution to the \texorpdfstring{$6j$}{6j}-contraction}

Shown below are the fragmentations of the trefoil knot that correspond to $4$-boundary wormholes contributing to the $6j$-contraction of 4 OPE coefficients. They are collectively described by the following diagram,
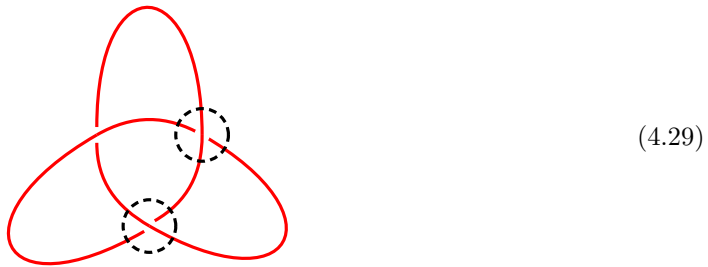
\begin{align}
     \vcenter{\hbox{
        %
}} 
\end{align}
where the dashed circle in the diagram indicates that the external Wilson line could stretch between any two adjacent segments of the knot which cross at the location where the circle is centered. This is a convenient notation to collectively describe fragmentations especially when there is more than one external Wilson line so we shall be employing this notation in the remaining examples. However, it is important to note that the $16$ fragmentations described by this figure are not all distinct.

The contribution from these fragmentations to the $6j$-contraction can be expressed collectively upto an overall phase depending on the spin of the operators as
\begin{equation}
    \overline{c_{12a}c_{34a}c_{23b}c_{41b}}\supset (\text{phase})\left |\sqrt{C_{12a}C_{34a}C_{23b}C_{14b}}\int dP \rho_0(P)e^{ 3\pi i P^2} 
\right|^2
\end{equation}

For illustration, we calculate below the partition function of one such fragmentation.

\begin{align}
    \begin{split}
        Z_V\left[\vcenter{\hbox{
        %
    \end{split}
\end{align}
The corresponding 4-boundary wormhole contributes to the 4-point non-Gaussianity,
\begin{equation}
    \overline{c_{12a}c_{34a}c_{23b}c_{41b}}\supset Z_{\text{grav}}\left[ %
\right|^2
\end{equation}
It is easy to see, for example, using the identities derived in Appendix \ref{Useful identities} that the partition functions on wormholes constructed using any of the other fragmentations only differ in an overall phase.

There is another class of fragmentations of the trefoil knot contributing to the $6j$-contraction. These involve adding an interaction vertex to the external Wilson line as shown in the diagrams below,
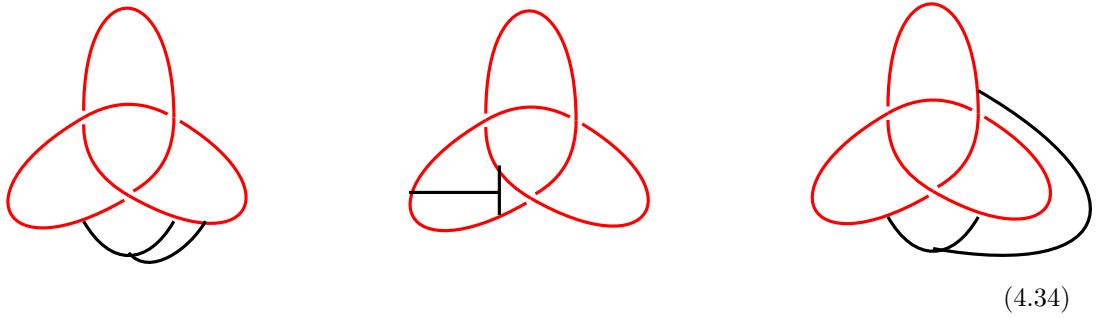
\begin{equation}
    \vcenter{\hbox{
        %
}}
\end{equation}
For illustration, we write down the VTQFT and the gravitational partition function corresponding to the knot diagram on the left in the above figure,
\begin{equation}
     Z_V\left[\vcenter{\hbox{
        %
\end{equation}
with the gravitational partition function on the corresponding wormhole given by
\begin{equation}
    Z_{\text{grav}}=(-1)^{\ell_2}\left | \sqrt{C_{12a}C_{34a}C_{23b}C_{14b}}%
 \right |^2
\end{equation}

\subsection{The figure-eight knot (4 crossings)}

\subsubsection{Contribution to the pillow contraction}

The following fragmentations of the figure-eight knot contribute to the pillow contraction of OPE coefficients,
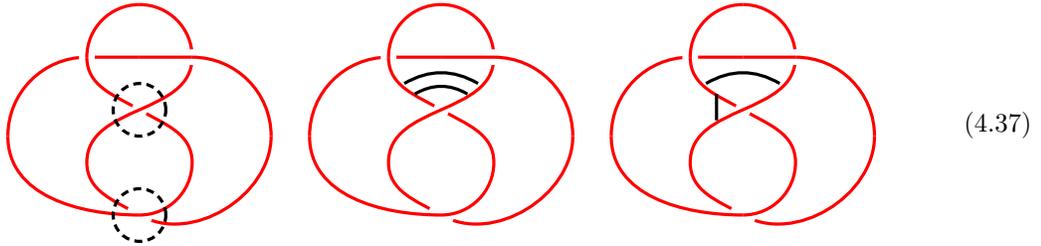
\begin{align}
    \vcenter{\hbox{ %
}}
 \end{align}
Even though these fragmentations correspond to different wormholes, the contribution of these diagrams to the pillow contraction can be expressed collectively as
\begin{align}
\begin{split}
   \overline{c_{12a}c_{13a}c_{24b}c_{34b}}\supset (\text{phase})\bigg|\sqrt{C_{12a}C_{13a}C_{24b}C_{34b}}\int & dP_s dP_t \rho_0(P_s) \rho_0(P_t) e^{2\pi i (P_s^2-P_t^2)}\\& \qquad \qquad \times %
\bigg |^2
     \end{split}
\end{align}

As an illustration, we compute the partition function on the following diagram belonging to the first class of fragmentations, 
\begin{align}
    \begin{split}
        Z_V\left[\vcenter{\hbox{ %
    \end{split}
\end{align}
Note that the effect of moving the external Wilson line labelled $b$ downwards through both the crossings only changes the phase in the VTQFT partition function,
\begin{equation}
            Z_V\left[\vcenter{\hbox{ %
}} \right]
\end{equation}
The 4-boundary wormhole corresponding to the wormhole on the right is drawn below,
\begin{equation} \label{fig8wormpillow}
    \overline{c_{12a}c_{1a3}c_{24b}c_{3b4}} \supset Z_{\text{grav}}\left[ %
\right |^2
\end{equation}
We can also check that the above result is consistent with the extended Gaussian ensemble of CFT$_2$ that incorporates the non-perturbative correction to the variance coming from the fragmentations of the figure-eight knot computed in section \ref{Figureeighttwopoint}. To this end, we compute the averaged product of two four-point functions $\overline{\langle \mathcal{O}_a\mathcal{O}_1\mathcal{O}_4\mathcal{O}_b \rangle \langle \mathcal{O}_a\mathcal{O}_1\mathcal{O}_4\mathcal{O}_b \rangle^*}$ in two different ways and show that they agree. First, we compute the average by expanding both the four-point functions using the $s$-channel,
\begin{align} \label{fig8schannel}
    \begin{split}
        &\overline{\langle \mathcal{O}_a\mathcal{O}_1\mathcal{O}_4\mathcal{O}_b \rangle \langle \mathcal{O}_a\mathcal{O}_1\mathcal{O}_4\mathcal{O}_b \rangle^*}\\ \\
        & =\sum_{2,3}\overline{c_{12a}c_{24b}c_{1a3}c_{3b4}}\left|  \vcenter{\hbox{
}} \bigg |^2
    \end{split}
\end{align}
where we substituted the gravity result (\ref{gravfig8pill}) for the non-Gaussianity.
Now, we compute the average by expanding both the four-point functions using the $t$-channel,
\begin{align}
    \begin{split}
        &\overline{\langle \mathcal{O}_a\mathcal{O}_1\mathcal{O}_4\mathcal{O}_b \rangle \langle \mathcal{O}_a\mathcal{O}_1\mathcal{O}_4\mathcal{O}_b \rangle^*}\\ \\
        & =\sum_{P,Q}\overline{c_{abP}c_{1P4}c^*_{abQ}c^*_{1Q4}}\left|  \vcenter{\hbox{%
}} \bigg |^2
    \end{split}
\end{align}
In the second line, we used (\ref{fig8var}) to evaluate the Gaussian average. In the third line, we expressed the $t$-channel blocks in terms of the $s$-channel blocks. In the last line, we used the pentagon identity to evaluate the integral over $P$ and the resulting expression matches with the result obtained using the $s$-channel expansion thereby showing that the gravity result (\ref{gravfig8pill}) is consistent with the extended Gaussian ensemble. Note that the averaged product of 4-point functions computed in (\ref{fig8schannel}) is also the gravitational partition function on a 2-boundary wormhole with the boundaries being 4-punctured spheres. The worldlines sourced by operators $\mathcal{O}_a$ and $\mathcal{O}_b$ are unknotted and go straight across the wormhole while the worldlines sourced by operators $\mathcal{O}_1$ and $\mathcal{O}_4$ are knotted into a figure-eight knot in exactly the same way as shown in (\ref{fig8wormpillow}).

\subsubsection{Contribution to the \texorpdfstring{$6j$}{6j}-contraction}

The following fragmentations of the figure-eight knot contribute to the $6j$-contraction of OPE coefficients,
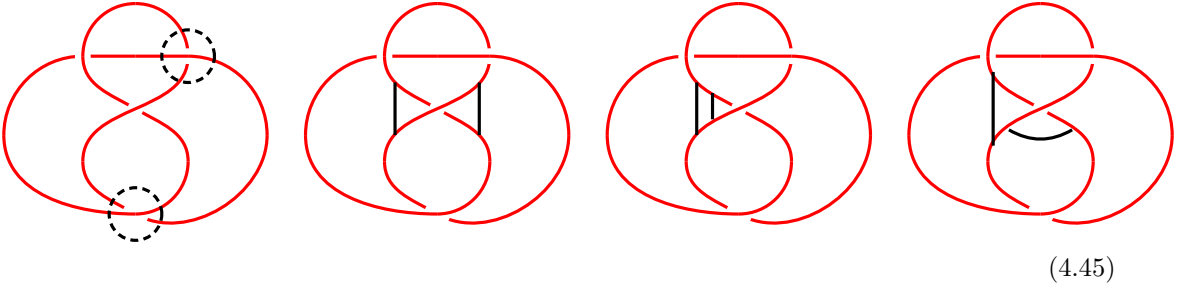
\begin{align}
    \vcenter{\hbox{ 
}} 
\end{align}
In the first diagram, the two external Wilson lines stretch across different crossings of the knot while in the rest of the diagrams, they stretch across the same crossing.
In addition to these diagrams, there are also diagrams with an interaction vertex on the external Wilson line. By application of the Wilson triangle identity, they can be reduced to diagrams with one Wilson line considered previously. The contribution from the fragmentations described by the first diagram in the above figure to the $6j$-contraction can each be represented as
\begin{equation}
\begin{split}
   \overline{c_{12a}c_{34a}c_{23b}c_{14b}} \supset (\text{phase})\bigg |\sqrt{C_{12a}C_{34a}C_{23b}C_{14b}}\int dP_s dP_t &\rho_0(P_s) \rho_0(P_t) e^{2\pi i (P_s^2-P_t^2)}\\& \quad \times
\bigg|^2
     \end{split}
\end{equation}
The contribution from the diagrams where both the Wilson lines stretch across the same crossing can be expressed as
\begin{equation}
\begin{split}
   \overline{c_{12a}c_{34a}c_{23b}c_{14b}} \supset (\text{phase})\bigg |\sqrt{C_{12a}C_{34a}C_{23b}C_{14b}}&\int dP_s  dP_t dP_d \rho_0(P_d) \rho_0(P_s) \rho_0(P_t) e^{i\pi (P_s^2+P_d^2-2P_t^2)}\\&  \times
   %
\bigg|^2
     \end{split}
\end{equation}

As an illustration, we compute the partition function on the following diagram belonging to the first class of fragmentations, 
\begin{align}
    \begin{split}
        Z_V\left[\vcenter{\hbox{ %
    \end{split}
\end{align}
The corresponding 4-boundary wormhole contributes to the 4-point non-Gaussianity,
\begin{equation}
    \overline{c_{12a}c_{34a}c_{2b3}c_{1b4}}\supset Z_{\text{grav}}\left[ %
\bigg|^2
     \end{split}
\end{equation}

\subsection{The Solomon's knot (4 crossings)}

Just like the Hopf link, the Solomon's knot being a two-component link only gives a contribution to the pillow contraction (and not the $6j$-contraction) when fragmented by two Wilson lines. The fragmentations contributing to the pillow contraction are sketched below,
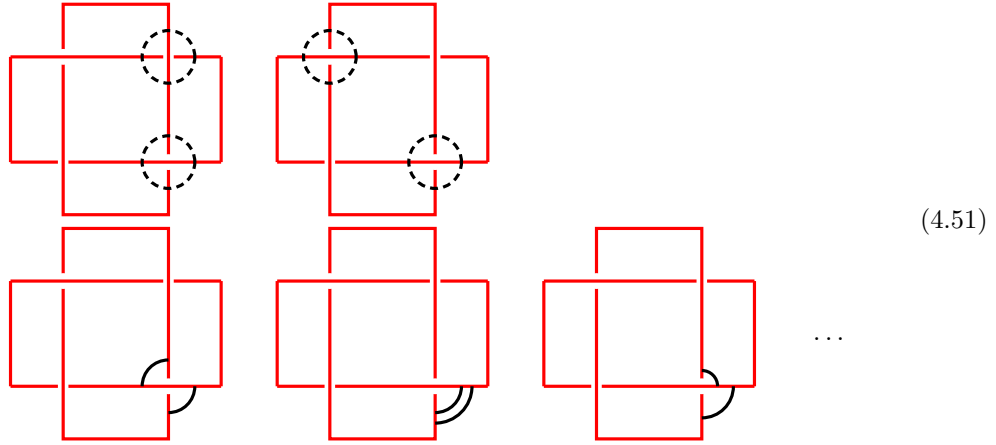
\begin{align} \label{Solpill}
\begin{split}
     &\vcenter{\hbox{%
}} \qquad \dots
    \end{split}
\end{align}
In the first line of diagrams, the external Wilson lines stretch across two different crossings of the link while in the second line of diagrams, the external Wilson lines stretch across the same crossing. The $\dots$ on the second line indicate there are more such diagrams. A convenient shorthand for diagrams on the second line could be
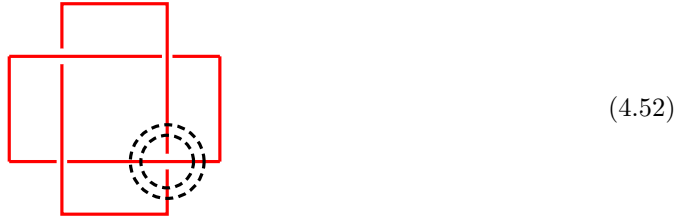
\begin{equation}
    \vcenter{\hbox{%
}}
\end{equation}
However, it is important to note that we only consider those diagrams which cannot be expressed in terms of a diagram on the first line of (\ref{Solpill}).
The contributions from the 4-boundary wormholes constructed from the first line of diagrams to the pillow contraction can be collectively expressed as 
\begin{equation} \label{solpil1}
    \overline{c_{12a}c_{12b}c_{34a}c_{34b}}\supset (\text{phase})\left |\sqrt{C_{12a}C_{34a}C_{12b}C_{34b}} \int dP \rho_0(P) e^{ 4\pi i P^2}%
\right |^2 
\end{equation}
while the contributions from the second line can be expressed as
\begin{align} \label{solpil2}
\begin{split}
    \overline{c_{12a}c_{12b}c_{34a}c_{34b}}\supset (\text{phase})\bigg |\sqrt{C_{12a}C_{34a}C_{12b}C_{34b}} \int & dPdP_d \rho_0(P) \rho_0(P_d)e^{ i\pi(-P^2+3P_d^2)}\\& \qquad \times%
\bigg |^2 
\end{split}
\end{align}
Note that unlike the previous examples in this section, there are two distinct contributions to the pillow contraction coming from fragmentations of the Solomon's knot. We will show below that this is tied to the fact the fragmentations of the Solomon's knot by a single Wilson line were giving two distinct contributions to the two-point non-Gaussianity as we observed in section \ref{Soltwopoint}.

For illustration, we explicitly write down the partition function on one of the fragmentations described by the first diagram on the first line of (\ref{Solpill}),
\begin{equation}
     Z_V\left[\quad \vcenter{\hbox{

\end{equation}
and the gravitational partition function on the corresponding 4-boundary wormhole is given by
\begin{equation}
    Z_{\text{grav}}=\left |\sqrt{C_{12a}C_{34a}C_{12b}C_{34b}} \int dP \rho_0(P) e^{4\pi i P^2}%
\right |^2 
\end{equation}

The partition function of the first diagram on the second line of (\ref{Solpill}) upon assigning momenta $P_1$ and $P_3$ to the fragments crossing each other three times; and momenta $P_2$ and $P_4$ to the fragments crossing each other once, is given by
\begin{align}
\begin{split}
    & Z_V\left[\vcenter{\hbox{%
\end{split}
\end{align}
The partition function of the other diagrams on the second line of (\ref{Solpill}) differs from the above result only by an overall phase.

Now, we turn toward verifying the consistency of the gravitational results (\ref{solpil1}) and (\ref{solpil2}) with the CFT$_2$ ensemble that incorporates the two-point non-Gaussianities computed from the Solomon's knot in section \ref{Soltwopoint}. It turns out that (\ref{solpil1}) is consistent with (\ref{solfrag1}) while (\ref{solpil2}) is consistent with (\ref{Solfrag2}). The calculation to show that (\ref{solpil1}) is consistent with (\ref{solfrag1}) is identical to the one presented in the last half of section \ref{hopfCFT} for the Hopf link, with the replacement of the braiding phase $e^{2\pi i P^2}\to e^{4\pi i P^2}$ in every step of that calculation. So, we skip the details. Now, we shall show that (\ref{solpil2}) is consistent with (\ref{Solfrag2}). To this end, we first compute the average below by expanding both the four-point functions using the $t$-channel,
\begin{align} \label{solschannel}
    \begin{split}
        &\overline{\langle \mathcal{O}_a \mathcal{O}_1 \mathcal{O}_1 \mathcal{O}_b \rangle \langle \mathcal{O}_a \mathcal{O}_3 \mathcal{O}_3 \mathcal{O}_b \rangle^*}\\ \\
        & =\sum_{2,4}\overline{c_{12a}c_{1b2}c_{3a4}c_{34b}}\left|  \vcenter{\hbox{
}}\bigg |^2
    \end{split}
\end{align}
Now, we compute the average by expanding both the four-point functions using the $t$-channel,
\begin{align}
    \begin{split}
        &\overline{\langle \mathcal{O}_a \mathcal{O}_1 \mathcal{O}_1 \mathcal{O}_b \rangle \langle \mathcal{O}_a \mathcal{O}_3 \mathcal{O}_3 \mathcal{O}_b \rangle^*}\\ \\
        & =\sum_{s,t}\overline{c_{abs}c_{11s}c^*_{abt}c^*_{33t}}\left|  \vcenter{\hbox{%
}}\bigg |^2
    \end{split}
\end{align}
In the last line, we used the pentagon identity to evaluate the integral over $P_s$,
\begin{equation}
    \int dP_s \rho_0(P_s) %
\end{equation}
Since the two ways of computing the average agree, the gravitational result (\ref{solpil2}) is consistent with (\ref{Solfrag2}).

\section{Six-point non-Gaussianities} \label{sixpointnongaussianity}

In this section, we discuss some structures of six-point non-Gaussianities corresponding to six-boundary wormholes constructed from fragmentations of knots and links by three Wilson lines. For simplicity, we illustrate using the fragmentations of the Hopf link by three Wilson lines. In the process, we describe some of the structures where the wormholes constructed from the fragmentations of Hopf link provides the leading contribution. At the end, we also discuss an interesting fragmentation of the trefoil knot by three Wilson lines which corresponds to a six-boundary wormhole whose partition function can be expressed as the integral of a product of three R-matrices. 

\subsection{The Hopf link with three Wilson lines}

We discuss fragmentations of the Hopf link by three Wilson lines. We illustrate using examples distinguished by the number of `self' and `cross' Wilson lines. A `self' Wilson line stretches across the same component of the link while a `cross' Wilson line joins the two components of the link. 

\subsubsection{All cross Wilson lines}

We compute the partition function of VTQFT on the Hopf link with three Wilson lines joining the two components. First, we compute the partition function by introducing an identity line between the two components followed by $\mathbb{F}$-move on the line and a pair of $\mathbb{B}$-moves to remove the crossings to get
\begin{align}
\begin{split}
    Z_V\left[\vcenter{\hbox{

    \end{split}
\end{align}
By embedding this network of Wilson lines in $S^3$ and excising balls around each junction, we can construct the following six-boundary wormhole which contributes to the six-point non-Gaussianity,
\begin{equation}
\overline{c_{12a}c_{2b3}c_{3c4}c_{45a}c_{5c6}c_{6b1}}\supset Z_{\text{grav}}\left[%
 \right]
\end{equation}
whose gravitational partition function is given by
\begin{multline}
    Z_{\text{grav}}= \left| \sqrt{C_{1a2}C_{2b3}C_{3c4}C_{4a5}C_{5c6}C_{6b1}}\int dP \rho_0(P) e^{2\pi i P^2}
    %
 \right|^2
\end{multline}
Notice that in the absence of braiding phases, the above integral expression simplifies by virtue of the pentagon identity to give the VTQFT partition function of the unlink joined by three Wilson lines,
\begin{equation}
     Z_V\left[\vcenter{\hbox{
        %
\end{equation}
This in turn corresponds to the following six-boundary wormhole which gives the leading contribution to the same six-point non-Gaussianity discussed already in \cite{Collier:2024mgv} where the contribution from this wormhole was also matched with the Gaussian CFT$_2$ ensemble,
\begin{equation}
\overline{c_{12a}c_{2b3}c_{3c4}c_{45a}c_{5c6}c_{6b1}}\supset Z_{\text{grav}}\left[%
\right]
\end{equation}

We can derive an alternate representation of the VTQFT partition function of the Hopf link network by Heegaard splitting along thrice-punctured tori. The resulting states on $\Sigma_{1,3}$ are
\begin{equation}
    \begin{split}
        & \bigg |\,\,\,\vcenter{\hbox{
        %
}}\bigg \rangle
    \end{split}
\end{equation}
After performing an $\mathbb{S}$-transformation on one of the blocks on the RHS to resolve the interlocking of the tori, we evaluate the inner product \cite{Collier_2023, Verlinde:1989ua} between the two states using
\vspace{1cm}
\begin{equation}
    \bigg \langle \vcenter{\hbox{
        %
}}\bigg \rangle = \frac{\delta(P_e-P_{e'})\delta(P_d-P_{d'})\delta(P-P_c)}{\rho_0(P_c)\rho_0(P_d)\rho_0(P_e)C_{a6d}C_{d3e}C_{ePP}}
\end{equation}
to get the following alternate representation of the VTQFT partition function on the Hopf link with three Wilson lines,
\begin{align}
    \begin{split}
         Z_V\left[\vcenter{\hbox{
        %
        \end{split}
\end{align}
In the second line, we used the relation between $\mathbb{S}$ and $\mathbb{F}$. To show that the two VTQFT expressions are equivalent, we applied the pentagon identity twice: first to evaluate the $P_e$ integral and then to evaluate the $P_d$ integral as we have shown in the last two lines of the above calculation. Conversely, requiring that the two VTQFT expressions are equivalent gives a three-dimensional derivation of a version of the pentagon identity which in our notation is expressed as
\begin{align}
    \int dP_d \mathbb{F}_{1d}%
 
\end{align}

\subsubsection{Two cross and one self Wilson lines}

With three Wilson lines, we can also construct a configuration where two of the Wilson lines join the two component circles of the Hopf link while the third line extends between the same circle. This configuration can be used to construct a new six-boundary wormhole that gives the leading contribution to a six-point non-Gaussianity described below.
\begin{align}
    \begin{split}
        Z_V\left[\vcenter{\hbox{
        %
    \end{split}
\end{align}
In the first line, we applied the Wilson triangle identity to reduce the network to the Hopf link with two Wilson lines whose partition function was computed earlier. This configuration corresponds to the following six-boundary wormhole which gives the leading contribution to the non-Gaussianity,
\begin{equation}
    \overline{c_{1a2}c_{2b3}c_{3b4}c_{45c}c_{5a6}c_{61c}}\supset Z_{\text{grav}}\left[\vcenter{\hbox{%
 \bigg |^2
    \end{split}
\end{align}

\subsubsection{One cross and two self Wilson lines}

Now, we construct a configuration where there is one Wilson line joining the two circles and two Wilson lines extending between the same respective circles. We can easily evaluate the VTQFT partition function on this configuration by applying the Wilson triangle identity twice which reduces this setup to the Hopf link joined by a single Wilson line.
\begin{align}
    \begin{split}
         Z_V\left[\vcenter{\hbox{
        %
    \end{split}
\end{align}
The six-boundary wormhole constructed from this configuration gives the leading contribution to the following six-point non-Gaussianity,
\begin{equation}
    \overline{c_{1a2}c_{3b4}c_{35c}c_{4b5}c_{6a1}c_{62c}}\supset Z_{\text{grav}}\left[%
\bigg |^2
    \end{split}
\end{align}

\subsubsection{Adding interactions between the external Wilson lines}

We now consider the Hopf link with two Wilson lines discussed earlier but now add the third Wilson line in a way such that it mediates an interaction between the two Wilson lines. The resulting configuration gives a leading contribution to a six-point non-Gaussianity discussed below. There are several ways to compute the partition function of the network using VTQFT. We first describe perhaps the simplest method which relates the network to the Hopf link with a single Wilson line. 
\begin{align}
    \begin{split}
         Z_V\left[\vcenter{\hbox{

    \end{split}
\end{align}
In the first line, we applied a $\mathbb{F}$-move on the interaction Wilson line labelled as $c$. In the second line, we first used the triangle identities in VTQFT to reduce to the network to a Hopf link with a single Wilson line which is known to evaluate to the modular $\mathbb{S}$-kernel. In the last line, we used the relation between the $\mathbb{S}$-kernel and the $\mathbb{F}$-kernel and exchanged the integration labels.

We can derive an alternate expression for the VTQFT partition function following our usual procedure of introducing a fictitious identity line and applying an $\mathbb{F}$-move on it. 
\begin{align}
    \begin{split}
        Z_V\left[\vcenter{\hbox{

    \end{split}   
\end{align}
In the second line, we applied an $\mathbb{F}$-move on the line labelled $4$. In the last line, we used the triangle identity a couple of times and simplified the resulting expression. One utility of this form of the expression is that upon removing the braiding phase from the integral, we easily recover the VTQFT partition function on the unlink joined by the same network of Wilson lines. To see this, we first evaluate the $P$ integral using the idempotency of the $6j$-symbol to get $\delta(P_1-P_3)$ upto normalisation factors and then evaluate the $P_d$ integral using the idempotency of $6j$-symbol to get $\delta(P_4-P_6)$ upto normalisation factors,
\begin{equation}
     Z_V\left[\vcenter{\hbox{
        \begin{tikzpicture}[scale=0.8]
        \draw[very thick, red] (-1.5,0) circle (1);
        \draw[very thick, red] (1.5,0) circle (1);
        \draw[very thick, red] ({-1.5+cos(45)},{sin(45)}) -- ({1.5-cos(45)},{sin(45)});
         \draw[very thick, red] ({-1.5+cos(45)},{-sin(45)}) -- ({1.5-cos(45)},{-sin(45)});
         \draw[very thick, red] (0,{sin(45)}) --(0,{-sin(45)});
         \fill[red] (0,{sin(45)}) circle (0.07);
         \fill[red] (0,{-sin(45)}) circle (0.07);
         \fill[red] ({-1.5+cos(45)},{sin(45)}) circle (0.07);
         \fill[red] ({1.5-cos(45)},{sin(45)}) circle (0.07);
         \fill[red] ({-1.5+cos(45)},{-sin(45)}) circle (0.07);
         \fill[red] ({1.5-cos(45)},{-sin(45)}) circle (0.07);
         \node[black] at (-2.2,0) {$a$};
        \node[black] at (2.2,0) {$b$};
        \node[black] at (-0.8,0) {$2$};
        \node[black] at (0.8,0) {$5$};
        \node[black] at (0.4,1) {$6$};
            \node[black] at (-0.4,1) {$1$};
            \node[black] at (-0.4,-1) {$3$};
            \node[black] at (0.4,-1) {$4$};
            \node[black] at (0.2,0) {$c$};
        \end{tikzpicture}  }}
    \right]=\frac{\delta(P_1-P_3)\delta(P_4-P_6)}{\rho_0(P_1)\rho_0(P_4)C_{1a2}C_{4b5}C_{6c1}}
\end{equation}
The above result can also be independently verified by applying the Wilson bubble identity twice to get the two $\delta$-functions with the right normalisation factors as shown in the RHS of the above expression. Requiring that the two expressions for the VTQFT partition functions on the Hopf link setup agree gives the following non-trivial integral identity involving crossing kernels,
\begin{align}
\begin{split}
    & \int dP dP_d \rho_0(P)\rho_0(P_d)e^{2\pi i P^2}

    \end{split}
\end{align}

We can now use the VTQFT partition function on the Hopf link configuration to express the gravitational partition on the following six-boundary wormhole which gives the leading contribution to the six-point non-Gaussianity,
\begin{equation}
    \overline{c_{1a2}c_{2a3}c_{34c}c_{4b5}c_{5b6}c_{61c}}\supset Z_{\text{grav}}\left[%
 \bigg|^2
    \end{split}
\end{align}

\subsection{The trefoil knot with three Wilson lines}

A large number of fragmentations of the trefoil knot by three Wilson lines corresponding to various structures of six-point non-Gaussianities can be readily written down. We will not exhaustively list all the fragmentations here. For example, there are 3 classes of fragmentations collectively described by the three diagrams below,
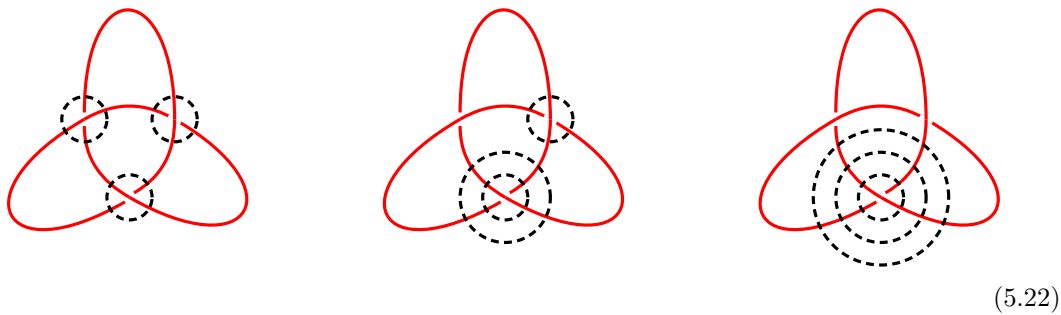
\begin{align}
     \vcenter{\hbox{
        %
}}
\end{align}
These are the diagrams when there are no interaction vertices on the external Wilson lines. The diagrams with vertices on the external Wilson lines can be broken down by application of triangle identity into diagrams with one or two Wilson lines considered in the earlier sections.

As an illustration, consider the knot diagram below belonging to the first class of fragmentations shown above, using which we can construct the following six-boundary wormhole discussed previously in \cite{Collier:2024mgv} that contributes to the six-point non-Gaussianity $\overline{c_{1a2}c_{2b3}c_{3c4}c_{4a5}c_{5b6}c_{6c1}}$,
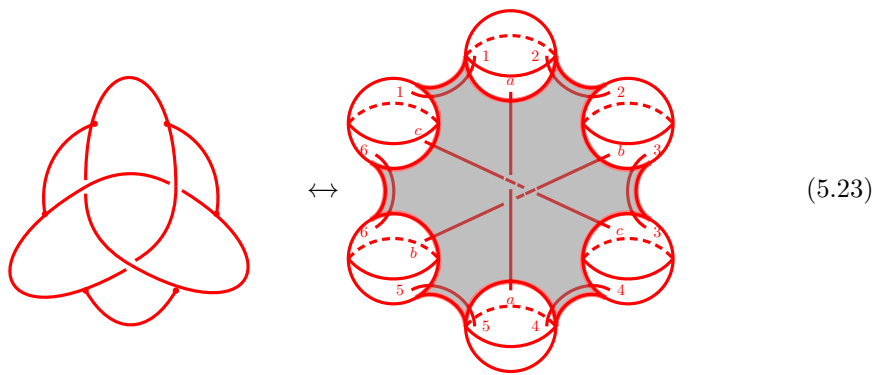
\begin{equation}
   \vcenter{\hbox{

\end{equation}
It is straightforward to calculate the partition function for this setup by applying $s-u$ crossing moves on the three external Wilson lines and check that the result agrees with the calculation done in \cite{Collier:2024mgv}. We shall omit the details here but would like to mention that the advantage of working with the knot diagram is that it is easy to see that the partition function is the integral of a  product of three R-matrices, so it can be readily matched with the prediction of the Gaussian ensemble of CFT$_2$ data by expanding the averaged product of three 4-point functions using the u-channel.
As an aside, it is interesting to note that the knot diagram shown above can also be obtained by turning three of the six crossings in the Borromean ring diagram into junctions. 

\section{Discussion}

In this paper, we have discussed non-perturbative Gaussian and non-Gaussian corrections to the OPE statistics using a framework that can generate a class of such non-perturbative corrections - \textit{Fragmentation of knots and links by Wilson lines}. We illustrated this idea by constructing multi-boundary wormholes from fragmentation diagrams of prime knots and links including non-hyperbolic ones with upto 5 crossings. These wormholes provide gravitational contributions to certain index contractions of OPE coefficients. In the process, we observed how the partition functions on wormholes related by different fragmentations of the same knot or link are closely related. Below, we make some interesting observations that could be potential future directions:

\begin{itemize}
   \item \textbf{Relation between fragmented knots and tangles}: It is interesting to note that there is a close relation between the fragmentation of knots and links discussed in this paper with $n$-tangles which are embeddings of $n$ Wilson lines in a 3-ball ending on $2n$ marked points on the boundary of the ball. Let us explain this with the help of an example that relates the fragmentation of a trefoil knot by a Wilson line to a rational 2-tangle in $\mathbb{H}_3$. The idea is to make the external Wilson line `heavy' in the sense that it is associated with an intermediate OPE channel in a conformal block decomposition. In effect, this turns the 2-boundary wormhole in (\ref{trefworm}) constructed from fragmentation of the trefoil knot into a hyperbolic ball with the pair of worldlines corresponding to the two knot fragments forming a rational 2-tangle (since the three crossings can be undone by a couple of monodromy transformations moving the marked points around each other on the boundary). The figure below depicts this relation between a fragmented trefoil knot (left) and a rational 2-tangle (right),
   \begin{align}
        \vcenter{\hbox{
}}
   \end{align}
   The black circle around two of the end-points of the Wilson lines in the figure on the right is drawn to indicate that the external Wilson line in the left figure now sets the monodromy around the non-contractible bulk cycle of the rational tangle. Integrating over the conformal weights around this cycle would give the 4-point Virasoro identity block acted on by an element of the Mapping Class Group of a 4-punctured sphere. Rational tangles like the one discussed above have appeared previously in the literature for example in \cite{Benjamin:2021wzr} where they were used to provide a geometrical interpretation for the individual terms of a modular sum of the Eisenstein series, appearing in the context of the averaged-Narain duality \cite{Afkhami-Jeddi:2020ezh, Maloney:2020nni}.

   \item \textbf{Organizing the sum over geometries}\footnote{I thank Scott Collier for discussions on this point.}: Maloney and Witten famously proposed a sum over geometries in 3d gravity by filling in the different cycles of a torus \cite{Maloney:2007ud}. In the dual CFT, this corresponds to a sum over modular images of the torus vacuum character and the resulting density of states can be conveniently expressed as a sum over PSL($2,\mathbb{Z}$)-crossing kernels \cite{Benjamin:2020mfz}. It would be interesting to consider an analogous sum over geometries where the solid tori are replaced by hyperbolic balls with a pair of worldlines anchored to the boundary which is a 4-punctured sphere. In the dual CFT, this would correspond to a sum over images of the 4-point Virasoro identity block under the Mapping Class Group of the 4-punctured sphere. This would extract a crossing-symmetric expression for the variance of the OPE coefficients written schematically below,
    \begin{equation}
        \overline{c_{ijk}c_{ijk}^*}=\sum_{\gamma \in \text{MCG}(\Sigma_{0,4})}\left|\mathbb{F}(\gamma)\right|^2
    \end{equation}
    where $\mathbb{F}(\gamma)$ is the Virasoro fusion kernel associated with the channel $\gamma \in \text{MCG}(\Sigma_{0,4})$\footnote{The Mapping Class Group of the 4-punctured sphere is closely related to the Mapping Class Group of the torus, $\text{MCG}(\Sigma_{0,4})=\text{PSL}(2,\mathbb{Z})\ltimes (\mathbb{Z}_2\times \mathbb{Z}_2)$. The two $\mathbb{Z}_2$ factors are associated with a pair of hyperelliptic involutions of $\Sigma_{0,4}$. See for example Chapter 2 of the book \cite{FarbMargalit+2012} for a detailed explanation.}. Such a sum over images for the variance of OPE coefficients was also mentioned in \cite{deBoer:2025rct} where they referred to it as the handlebody part of the sum over geometries contributing to the variance. They also mention that there are non-handlebody corrections to this sum. 

    In the light of the present paper where we computed non-perturbative corrections to the variance from fragmentations of knots, it would be interesting to understand which of the fragmented knots contribute to the handlebody part of the sum and which of them give non-handlebody corrections. The relation between fragmented knots and tangles that we discussed in the previous point suggests that: \textit{Fragmented knots that correspond to rational 2-tangles contribute to the handlebody part of the sum over geometries and the fragmented knots that correspond to non-rational 2-tangles give non-handlebody corrections.}

    \item \textbf{Self-energy divergences from hyperbolic knots}: In all the examples considered in this paper, the external Wilson line was stretched across atleast one crossing of the knot. But we could consider examples like the ones shown below where the external Wilson line is `contractible' (does not stretch across any crossing of the knot),
    \begin{equation}
         \vcenter{\hbox{
}}
    \end{equation} 
    Resolving the Wilson bubbles between the external Wilson line and the knot fragment gives a divergence proportional to $\delta(0)$. After this, it is not clear how to proceed with the trefoil knot example as it is not hyperbolic. But for the figure-eight knot, the divergence would just be multiplied by the VTQFT partition function on the knot complement. The two-boundary wormhole constructed from such a fragmentation of the figure-eight knot has one of the worldlines knotted with itself into a figure-eight knot, and the other two worldlines are unknotted. From the explanation above, it appears that such a wormhole amplitude computed using VTQFT is divergent. It would be necessary to understand if such divergences coming from knots on individual worldlines can simply be renormalised away.

    \item \textbf{External Wilson lines crossing knot fragments}: In this paper, we have considered examples where the external Wilson line does not cross the fragments of knots or links. As a consequence, the two-boundary wormholes that we constructed only had two of the worldlines knotted with each other. But it would be interesting to consider fragmentations where the external Wilson line crosses the fragments. These would correspond to wormholes where all three worldlines are knotted in the bulk. An example of such a fragmentation of the figure-eight knot is sketched below,
    \begin{equation}
        \vcenter
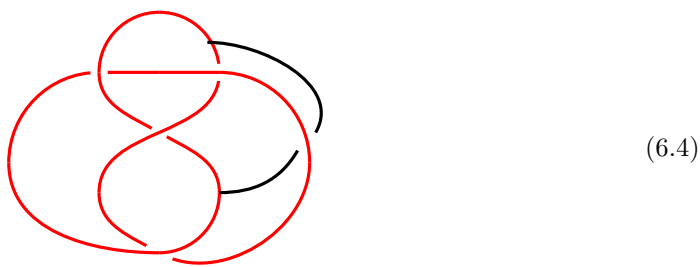
{\hbox{ \begin{tikzpicture} [scale=0.8]      
        \draw[very thick, out = 90, in = 180, red] (-5/2,-3/2) to (-1,0);
        \draw[very thick, out = 0, in = 180, red] (-1,0) to (0,0);
        \draw[fill=white,draw=white] (-1,0) circle (1/10);
        \draw[very thick, out = 180, in = 90, red] (0,1) to (-1,0);
\draw[very thick, out = 270, in = 150, red] (-1,0) to (0,-1);
        \draw[very thick, out = 330, in = 90, red] (0,-1) to (1,-2);
        \draw[fill=white,draw=white] (0,-1) circle (1/10);
\draw[very thick, out = 90, in = 270, red] (-1,-2) to (1,0);
        \draw[very thick, out = 90, in = 0, red] (1,0) to (0,1);
        \draw[fill=white,draw=white] (1,0) circle (1/10);
        \draw[very thick, out = 0, in = 180, red] (0,0) to (1,0);
        \draw[very thick, out = 0, in = 90, red] (1,0) to (5/2,-3/2);
        \draw[very thick, out = 270, in = 330, red] (5/2,-3/2) to (0,-3);
\draw[very thick, out = 150, in = 270, red] (0,-3) to (-1,-2);
        \draw[fill=white,draw=white] (0,-3) circle (1/5);
\draw[very thick, out = 270, in= 0, red] (1,-2) to (0,-3);
        \draw[very thick, out = 180, in = 270, red] (0,-3) to (-5/2,-3/2);
        \draw[very thick, black, out=0, in=60, looseness=1] (.8,.5) to (2.6,-1);
         \draw[very thick, black, out=-120, in=0, looseness=1] (2.3,-1.3) to (1,-2);
\end{tikzpicture}}}
    \end{equation}

    \item \textbf{Fragmentations of composite knots}: In this paper, we have only discussed fragmentations of prime knots and links with upto five crossings. It would be interesting to also study fragmentations of composite knots obtained by taking a connected sum of prime knots. Since the simplest non-trivial prime knot is the trefoil knot, the simplest composite knots are obtained from a connected sum of two trefoil knots. There are two such composite knots: the granny knot denoted as $3_1 \# 3_1$ is the connected sum of two trefoil knots of the same chirality; and the square knot denoted as $3_1 \# 3_1^*$ is the connected sum of two trefoil knots of opposite chiralities. Both these composite knots have 6 crossings.

    \item \textbf{Fragmentations of higher-genus handlebody knots}: In this paper, we described fragmentations of genus-1 handlebody knots (knotted embeddings of solid tori in $S^3$) by external Wilson lines. It would be interesting to extend these constructions to the knotted embeddings of higher-genus handlebodies in $S^3$. See for example \cite{doi:10.1142/S0218216511009893} for a list of genus-2 handlebody knots with upto six crossings. Even without the addition of external Wilson lines, they compute non-perturbative corrections to the variance or to the two-point non-Gaussianity \cite{Scottetal} depending on the specific knot. It would be interesting to compare the partition functions of the genus-2 handlebody knots to the partition functions of the fragmentations of genus-1 handlebody knots considered in this paper. In addition, one could also study fragmentations of higher-genus handlebody knots by external Wilson lines which contribute to non-perturbative corrections to higher moments of the OPE data.
\end{itemize}

\ \\ 

\ \\

\noindent \textbf{Acknowledgments}\ \\

I thank Scott Collier, Lorenz Eberhardt, Tom Hartman, Diego Liska, Eric Perlmutter, Boris Post, Ioannis Tsiaries, Gabriele Ubaldo and Diandian Wang for discussions on topics related to this work. I also thank Scott Collier for valuable comments on a draft of this work. This work was supported in part by the Boochever Fellowship and in part by the NSF grant PHY-2014071.

\appendix

\section{VTQFT identities and crossing kernels} \label{VTQFT identities}

In this appendix, we list some useful VTQFT identities and consistency conditions obeyed by the crossing kernels which were used in the main text. For a recent comprehensive review about bootstrapping crossing kernels from the Moore-Seiberg consistency conditions, refer to \cite{Eberhardt:2023mrq}. 

\begin{itemize}
    \item Normalisation of vertices:
    \begin{equation}
        \vcenter{\hbox{
}}
    \end{equation}
    \item Wilson bubble identity: 
    \begin{equation}
        \vcenter{\hbox{%
}}
    \end{equation}
    \item Wilson triangle identity:
    \begin{equation}
        \vcenter{\hbox{%
    \end{equation}
    The $6j$-symbol in the Racah-Wigner normalisation manifests tetrahedral symmetry: Invariant under exchange of any two columns and invariant under exchange of any two elements of the first row with corresponding elements of the second row.
    \item Braiding phases:
    \begin{equation} \label{braid}
        \vcenter{\hbox{%
}}
    \end{equation}
    \\
    \\
    where the braiding phase is given by $\mathbb{B}^{a,b}_c\equiv e^{i\pi(\Delta_c-\Delta_a-\Delta_b)}$. When the pattern of over- and under-crossings is reversed, the braiding phase flips.
    \item The Fusion kernel $\mathbb{F}$:
    \begin{equation}
        \vcenter{\hbox{%
=\delta(P_a-P_b)
\end{equation}
    \item Relation between fusion kernel $\mathbb{F}$ and the $6j$-symbol in Racah-Wigner normalisation:
    \begin{equation}
        \mathbb{F}_{3d}%
}}
    \end{equation}\\
    The kernel in the above $s-u$ crossing transformation is called the R-matrix.
    When the pattern of under- and over-crossings is reversed, the phase in the R-matrix flips. It is important to note that the convention used for the braiding phase in (\ref{braid}) and the one above for the R-matrix must be consistent with each other.
    \item Symmetry of $\mathbb{S}$:
    \begin{equation}
        \frac{\mathbb{S}_{ab}[c]}{\rho_0(P_b)C_{bbc}}=\frac{\mathbb{S}_{ba}[c]}{\rho_0(P_a)C_{aac}}
    \end{equation}
    \item Relation between modular-$\mathbb{S}$ kernel and fusion kernel $\mathbb{F}$:
    \begin{align}
        \mathbb{S}_{ab}[c]=&\int dP \rho_0(P_b) \frac{C_{bbc}}{C_{abP}}e^{i\pi(2\Delta_P+\Delta_c-2\Delta_a-2\Delta_b)}\mathbb{F}_{cP}

    \end{align}
    The two expressions are equivalent since $e^{-\frac{i\pi}{2}\Delta_c}\mathbb{S}_{ab}[c]$ is real.
    \item Relation between $\mathbb{F}$ and $\mathbb{F}^{-1}$:
    \begin{equation}
     \mathbb{F}_{bP}%
\end{equation}
    \item The pantagon identity:
    \begin{align}
    \int dP_d \mathbb{F}_{1d}%
 
\end{align}
This is not the most general form of the pentagon identity but suffices for applications in this paper.
\item The hexagon identity:
\begin{equation}
    \int dP e^{\pm i\pi(\sum_{i=1}^4 \Delta_i -\Delta_s-\Delta_t-\Delta_P)}\mathbb{F}_{sP}%
\end{equation}
\end{itemize}

\section{Useful identities involving subdiagrams of knot fragmentations} \label{Useful identities}

We make a note here of some useful subdiagrams that show up when we compute the partition functions of knot or link fragmentations. These identities involve the different ways in which the external Wilson line can stretch across a crossing. We have used identities listed in Appendix \ref{VTQFT identities} to arrive at the expressions on the RHS below,
\begin{enumerate}
    \item \begin{align}
     \vcenter{\hbox{%
}}
\end{align}
\end{enumerate}
Observe that the three diagrams give the same expression upto an overall phase. This observation helps understand why different fragmentations of the same knot or link give similar VTQFT partition functions.

\section{The three-twist knot (5 crossings)} \label{Three-twist}

The three-twist knot denoted $5_2$ in the Alexander-Briggs notation is a hyperbolic knot. It is described by a Wilson loop with $5$ crossings,
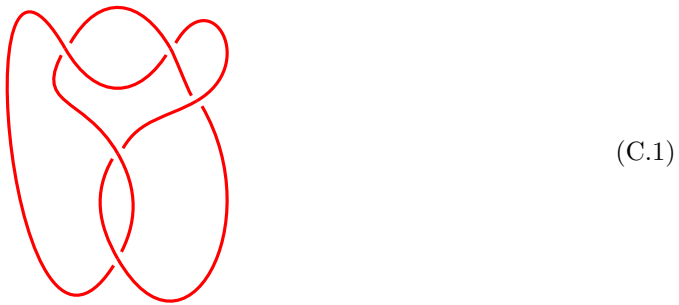
\begin{equation}
    \vcenter{\hbox{%
}}
\end{equation}
The partition function on its complement in $S^3$ can be readily computed using VTQFT,
\begin{equation} \label{52part}
    Z_V[5_2]=e^{-2\pi i \Delta_0}\int dP_s dP_t  \rho_0(P_s)\rho_0(P_t)e^{\pi i (3\Delta_t-2\Delta_s)}%
\end{equation}
Here $P_0$ is the Liouville momentum of the Wilson loop usually tuned to the threshold value corresponding to the cusp $\Delta_0=\frac{c-1}{24}$. But we can let it take a general value. We derive the above expression in the section below, where we discuss the fragmentation of $5_2$ knot by a Wilson line. The complement of the $5_2$ knot is a hyperbolic 3-manifold with a finite volume, $\text{vol}(5_2)=2.82812$ (upto 5 decimal places). This can be calculated numerically, for example, by using the package $\texttt{SnapPy}$ by triangulating the complement using hyperbolic tetrahedra. The package also helps visualize the topology of the knot complement. It would be interesting to take the semiclassical limit of the VTQFT partition function and check that it reproduces this volume, which would provide a check of the volume conjecture,
\begin{equation}
    |Z_V[5_2]|=e^{-\frac{c}{12\pi}\text{vol}(5_2)}
\end{equation}
The absolute value is necessary since the $5_2$ knot is chiral, so the partition function is not expected to be real.
We will not check the volume conjecture in this paper as it is not relevant to the main topic of non-Gaussianities. As an aside, it would also be interesting to perform Dehn surgery on the $5_2$-knot complement, which generates a family of closed hyperbolic 3-manifolds labeled by a pair of coprime integers corresponding to the slope of the meridian cycle of the solid torus being glued in, and compare the semiclassical limits of the VTQFT partition functions on these manifolds to the known expressions for their volumes easily calculable numerically using \texttt{SnapPy}. For the figure-eight knot, this was done in \cite{Collier:2024mgv}.

\subsection{Fragmentation by one Wilson line}

Now, we discuss the fragmentation of the $5_2$ knot by an external Wilson line. Like in the examples presented in the main text, there are various fragmentations of the $5_2$ knot by a Wilson line that differ only in the pattern of crossings between the resulting fragments. But unlike the trefoil knot example or the figure-eight knot example where the external Wilson line stretches across a single crossing, the external Wilson line could also stretch across two crossings of the $5_2$ knot. We discuss one such fragmentation below. First, we compute the partition function on a fragmentation where the Wilson line stretches across a single crossing of the knot with the resulting two fragments crossing each other $5$ times,
\begin{align}
    \begin{split}
        Z_V\left[
    \vcenter{\hbox{

    \end{split}
\end{align}
In the last line, we used the hexagon identity to evaluate the integral over $P_u$,
\begin{equation}
    \int dP_u \rho_0(P_u) (\mathbb{B}^{1,2}_u)^{-1}%
=\mathbb{B}^{1,2}_t
\end{equation}
Taking the identity limit of the external Wilson line $P_a \to \frac{iQ}{2}$, we recover the partition function of the $5_2$ knot quoted earlier in (\ref{52part}).
The gravitational partition function on the two-boundary wormhole constructed from this fragmentation of the $5_2$ knot is 
\begin{align}
\begin{split}
    Z_{\text{grav}}=(-1)^{\ell_1+\ell_2}|C_{12a}|^2\bigg|&\int dP_s dP_t \rho_0(P_s)\rho_0(P_t)e^{\pi i (3P_t^2-2P_s^2)}%
\bigg|^2
        \end{split}
\end{align}
The wormhole obeys the boundary conditions corresponding to $\overline{c_{12a}^2}$, but if we instead express it as a contribution to $\overline{|c_{12a}|^2}$, we get
\begin{equation} \label{var52}
    \overline{|c_{12a}|^2} \supset (-1)^{\ell_a}|C_{12a}|^2\bigg|\int dP_s dP_t \rho_0(P_s)\rho_0(P_t)e^{\pi i (3P_t^2-2P_s^2)}%
\bigg|^2
\end{equation}
An interesting fragmentation in which the external Wilson line stretches across two crossings of the knot is sketched below. This is the first time we have encountered such a situation since it requires the knot to have atleast 5 crossings,
\begin{align}
    \begin{split}
       & Z_V\left[\vcenter{\hbox{%
    \end{split}
\end{align}
Again, in the identity limit, $P_a \to \frac{iQ}{2}$, the above expression reduces to (\ref{52part}) since the $6j$-symbol within the parentheses becomes a $\delta$-function $\delta(P_t-P_d)$ in this limit.
Its contribution to the variance can be expressed as
\begin{equation}
     \overline{|c_{12a}|^2} \supset |C_{12a}|^2 \left|\int dP_s dP_t dP_d \rho_0(P_s)\rho_0(P_t)\rho_0(P_d)e^{i\pi(2P_t^2-2P_s^2+P_d^2)}
        \left(%
\right |^2
\end{equation}
Therefore, we have shown that there are two distinct contributions to the variance from the $5_2$ knot depending on whether the external Wilson line stretches across one or two crossings of the knot.

\subsection{Fragmentation by two Wilson lines}

Now, we discuss fragmentations of the $5_2$ knot by two Wilson lines. Like in the examples presented in the main text, there are fragmentations corresponding to four-boundary wormholes that contribute to the pillow contraction or the $6j$-contraction of OPE coefficients.

\subsubsection{Contribution to the pillow contraction}

We illustrate the contribution of the $5_2$ knot to the pillow contraction of four OPE coefficients using the fragmentation shown below,
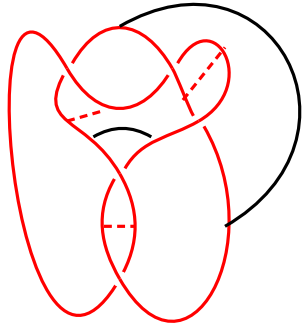
\begin{equation}
     \vcenter{\hbox{
}}
\end{equation}
Here, the dashed lines are identity insertions while the solid black lines are the external Wilson lines. The VTQFT partition function on this fragmentation takes the form,
\begin{align}
    \begin{split}
        Z_V=& \frac{1}{\sqrt{C_{12a}C_{13a}C_{24b}C_{34b}}}\int dP_s dP_t dP_d \rho_0(P_s)\rho_0(P_t)\rho_0(P_d)(\mathbb{B}^{1,3}_s)^{-2}(\mathbb{B}^{1,4}_t)^{2}\mathbb{B}^{1,2}_d\\ & \qquad \qquad \qquad \qquad \times 
        %
    \end{split}
\end{align}
The gravitational partition function on the corresponding 4-boundary wormhole is given by
\begin{align}
    \begin{split}
        Z_{\text{grav}}=& (-1)^{\ell_1+\ell_2}\bigg |\sqrt{C_{12a}C_{13a}C_{24b}C_{34b}}\int dP_s dP_t dP_d \rho_0(P_s)\rho_0(P_t)\rho_0(P_d)e^{i\pi(2P_t^2+P_d^2-2P_s^2)}\\ & \qquad \qquad \qquad \qquad \times 
        %
\bigg |^2
    \end{split}
\end{align}

\subsubsection{Contribution to the \texorpdfstring{$6j$}{6j}-contraction}

There are several fragmentations which contribute to the $6j$-contraction. We illustrate using one such fragmentation,
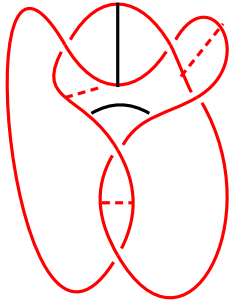
\begin{equation}
     \vcenter{\hbox{%
}}
\end{equation}
The solid black lines are the external Wilson lines and the dashed lines are the identity lines. The VTQFT partition function on this fragmentation takes the form,
\begin{align}
\begin{split}
    Z_V=\frac{1}{\sqrt{C_{12a}C_{34a}C_{23b}C_{41b}}}\int & dP_sdP_t \rho_0(P_s)\rho_0(P_t)(\mathbb{B}^{1,3}_s)^{-2}(\mathbb{B}^{3,4}_t)^{2}\mathbb{B}^{1,2}_t\\
    & \qquad \qquad \times %
    \end{split}
\end{align}
and the gravitational partition function on the corresponding 4-boundary wormhole is given by
\begin{align}
\begin{split}
    Z_{\text{grav}}=(-1)^{\ell_1+\ell_2}\bigg|\sqrt{C_{12a}C_{34a}C_{23b}C_{41b}}\int & dP_sdP_t \rho_0(P_s)\rho_0(P_t)e^{i\pi(3P_t^2-2P_s^2)}\\
    & \qquad \qquad \times %
    \bigg |^2
    \end{split}
\end{align}

\section{The Cinquefoil knot (5 crossings)} \label{Cinquefoil}

Apart from the three-twist knot, there is one other prime knot with 5 crossings called the Cinquefoil knot and is denoted $5_1$ in the Alexander-Briggs notation,
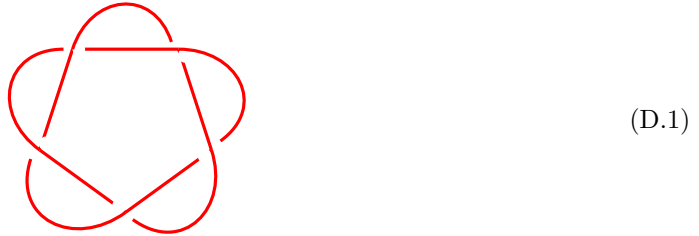
\begin{equation}
    \vcenter{\hbox{%
}}
\end{equation}
Unlike the three-twist knot, the cinquefoil knot is not hyperbolic. So, we don't expect to compute its partition function using VTQFT. However, just like the trefoil knot and the Solomon's knot examples in the main text, we can compute the partition function once we add a Wilson line fragmenting the $5_1$ knot. For illustration, we compute the partition function on the following fragmentation where the two fragments cross each other 5 times,
\begin{align}
    \begin{split}
        Z_V\left[\vcenter{\hbox{%
    \end{split}
\end{align}
We see that in the identity limit $P_a \to \frac{iQ}{2}$, the integral becomes $\int dP \rho_0(P) e^{5\pi iP^2}$ which does not admit a sensible saddle and is an evidence of the non-hyperbolicity of the knot complement. The wormhole constructed from this fragmentation contributes to $\overline{c_{12a}^2}$. If we instead express the result as a contribution to $\overline{|c_{12a}|^2}$, we get the following correction to the variance,
\begin{equation} \label{var51}
    \overline{|c_{12a}|^2} \supset (-1)^{\ell_a}|C_{12a}|^2\left|\int dP \rho_0(P)e^{5\pi i P^2}\begin{Bmatrix}
            1 & 2 & P\\
            1 & 2 & a
        \end{Bmatrix} \right|^2
\end{equation}
It would be interesting to compare the contribution (\ref{var51}) to the corresponding contribution from the three-twist knot (\ref{var52}) in appropriate semiclassical limits to check which of the two wormholes dominates the gravitational path integral. The correction to the variance in (\ref{var51}) holds whenever the external Wilson line stretches across a single crossing of the knot but when it stretches across two crossings, the result involves an additional integral with an additional $6j$-symbol,
    \begin{align}
    \begin{split}
        Z_V\left[\vcenter{\hbox{
 \right )^2
    \end{split}
\end{align}
with the correction to the variance given by
\begin{equation}
    \overline{|c_{12a}|^2} \supset |C_{12a}|^2 \left |\int dP_s dP_t \rho_0(P_s)\rho_0(P_t) e^{i\pi(3P_t^2+2P_s^2)}\left(%
 \right )^2\right |^2
\end{equation}

\bibliographystyle{ourbst}
\bibliography{ref.bib}

\end{document}